\documentclass[fleqn,usenatbib]{mnras}

\usepackage{newtxtext,newtxmath}
\usepackage{ulem}
\usepackage[T1]{fontenc}

\usepackage{soul}

\usepackage{multirow}
\usepackage{color}

\DeclareRobustCommand{\VAN}[3]{#2}
\let\VANthebibliography\thebibliography
\def\thebibliography{\DeclareRobustCommand{\VAN}[3]{##3}\VANthebibliography}

\usepackage{graphicx}	
\usepackage{amsmath}	

\title[Varying subpulse drifting of PSR J1857+0057]
{Quasi-regular variations of subpulse drifting for PSR J1857+0057}

\author[Y. Yan et al.]{
Yi Yan,$^{1,2}$
J.~L. Han,$^{1,2,3}$\thanks{E-mail: hjl@bao.ac.cn (JLH)}
C. Wang$^{1,2,3}$\thanks{E-mail: wangchen@bao.ac.cn (WC)}
and P.~F. Wang$^{1,2,3}$
\\
$^{1}$National Astronomical Observatories, Chinese Academy of Sciences, Beijing 100101, China\\
$^{2}$School of Astronomy, University of Chinese Academy of Sciences, Beijing 100049, China\\
$^{3}$CAS Key laboratory of FAST, National Astronomical Observatories, Chinese Academy of Sciences, Beijing 100101, China
}

\date{Accepted XXX. Received YYY; in original form ZZZ}

\pubyear{2015}

\begin{document}
\label{firstpage}
\pagerange{\pageref{firstpage}--\pageref{lastpage}}
\maketitle

\begin{abstract}
During observations of the Galactic Plane Pulsar Snapshot (GPPS) survey by the Five-hundred-meter Aperture Spherical radio Telescope (FAST), varying subpulses drifting of PSR J1857+0057 is detected. The following-up observation confirms the quasi-regularly changes of the drifting rate about every 50 periods. We determine the drift rate $D$ through a linear fit to the pulse-central longitudes of subpulses in a drifting band, and determine $P_3$ from the cross-points of two fitted lines at the zero longitude for two neighbored drifting bands. The low frequency modulation of about every 50 periods is found on variations of not only pulse intensity but also drift parameters. In most of low frequency modulation cycles, the integrated pulse intensity $I$ and the absolute drift rate $|D|$ tend to increase first and then decrease, and the drifting periodicity $P_3$ varies just in the opposite. 
In addition, the phase-forward intensity-enhancement is observed in many modulation cycles.
Based on our polarization data, the average PA curve for pulses with a smaller $|D|$ and larger $P_3$ is slightly steep in the leading edge of pulse profile compared with that of the fully averaged profile. 
\end{abstract}

\begin{keywords}
polarization -- pulsars: individual: PSR J1857+0057
\end{keywords}



\section{Introduction}           
\label{sect:intro}

Though mean profiles of pulsars are stable, individual pulses show diverse behaviours in single pulse sequences, such as nulling, mode changing and subpulse drifting. Nulling is the phenomenon that emission cannot be detected for some periods, and then returns to the normal pulse emission state. 
Mode changing is the transition between two or more quasi steady radiation states. Individual 
pulses often consist of many subpulses, and for some pulsars, subpulses drift continually for some periods inside the mean-pulse window, which is known as subpulse drifting \citep[e.g., ][]{Drake1968,Ruderman1975,Huguenin1970,Esamdin2005,Weltevrede2006}. 
If the drifting pattern of a pulsar is stable, then 
the drifting periodicity $P_3$ is defined as the period numbers for a drifting subpulse back to a given longitude after some periods; the phase interval $P_2$ is defined as the longitude difference in a period for two adjacent drift subpulses; and the drift rate $D$ is then defined as the phase shift of subpulses per period. In a stable drifting case, $D=P_2/P_3$.

Traditionally, the subpulse drifting has been explained by the rotating carousel of sub-beams due to the $E \times B$ drift in the inner acceleration region \citep{Ruderman1975}. As a result, the existence of the tertiary modulation related to the carousel rotation periodicity is expected. 
The low frequency feature in addition to the primary modulation in fluctuation spectra has previously been detected for PSRs B0943+10 \citep{Deshpande2001}, B1857-26 \citep{Mitra2008} and B1237+25 \citep{Maan2014}, indicating the carousel circulation modulation. 

Since the discovery of drifting phenomenon by \citet{Drake1968}, various unusual behaviors of subpulse drifting have been reported, challenging the traditional carousel circulation model. 
 \citet{Drake1968} first reported the drift rate variation of PSR B2016+28.   %
Some pulsars can have two or more drifting modes, such as PSRs J1727-2739 \citep{Wen2016_J1727m2739,Rukiye2021}, J1946+1805 \citep{Deich1986,Kloumann2010}, J1921+1948 \citep{Hankins1987,Rankin2013}, J1822-2256 \citep{Basu2018a}, J2006-0807 \citep{Basu2019}, B2303+30 \citep{Redman2005} and J2321+6024 \citep{Wright1981,Rahaman2021}.
A number of pulsars have subpulse drift rates changed continually. For example, PSR B0818$-$41 frequently has a drift rate changed and it is even related to nulling phenomenon  \citep{Bhattacharyya2009, Bhattacharyya2010}. PSR J0034-0721 has three drift modes \citep{Huguenin1970,Vivekanand1997,McSweeney2017}, with the drift rate changed gradually within each mode but abruptly between different modes.
`Drift reversal' behavior that subpulses change drift direction with time has been found in B0826$-$34 \citep{Biggs1985,Gupta2004,Esamdin2005} and J1750$-$3503 \citep{Szary2022}.
The driftings of different profile components for a given pulsar can have opposite drift directions, which has been called `bi-drifting', found in PSRs J0815+0939 \citep{Champion2005}, J1034$-$3224 \citep{Basu2018b}, B1839$-$04 \citep{Weltevrede2016} and J2006$-$0807 \citep{Basu2019}.

PSR J1857+0057 (B1854+00) was discovered by \citet{Mohanty1983} in a search of pulsars near supernova remnants at 327 MHz by using the Ooty Radio Telescope. It has a rotation period of $P=0.35693$~s and dispersion measure of 82.39~cm$^{-3}$pc \citep{Hobbs2004}. The pulsar was classified as the "conal single" category \citep{Rankin1983} at 1418 MHz because of moderate linear and circular polarization and edge depolarization \citep{Weisberg1999}, and its wider profile was reported at 430 MHz by \citet{Weisberg2004}. Our observations by the Five-hundred-meter Aperture Spherical radio Telescope (FAST) for the first time reveal its peculiar drifting phenomenon, especially the quasi-regular low frequency modulation on drifting. In Section~\ref{sect:ObsDataPro}, we briefly present observations and data processing. In Section~\ref{sect:LowFreModu}, we analyze subpulse drifting and the low frequency modulation. Discussion and conclusions are given in Section~\ref{sect:Discussion}.

\begin{figure}
\centering
\setlength\tabcolsep{0pt}
\includegraphics[width=0.45\textwidth]{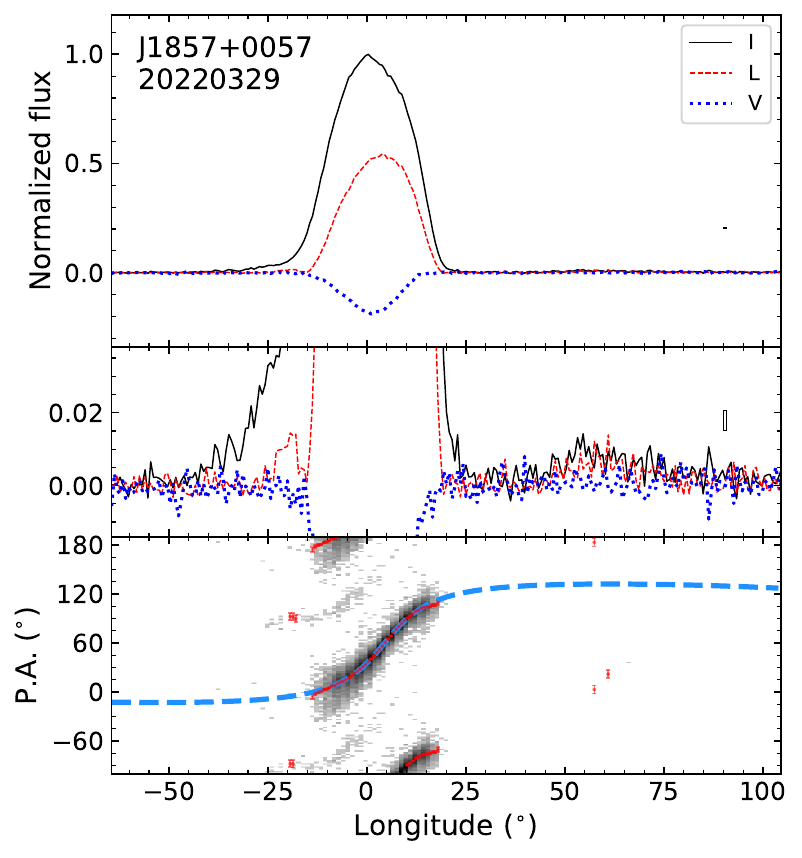}
\caption{The mean polarization profile of PSR J1857+0057 and linear polarization position angle (PA) distribution we observed at 1.25~GHz in the second FAST observation session on March 29, 2022 (20220329), with a time resolution of 512 bins in a period. {\it The top sub-panel} shows the total intensity $I$ (solid line), linear polarization $L$ (dashed line) and circular polarization $V$ (dotted line), together with a small box for one bin width and height of $\pm1\sigma$ of $I$,  and {\it the central sub-panel} is the zoomed view of the intensity profiles for illustration of the week extended emission. {\it The bottom sub-panel} shows the PA distribution (grey dots) for the bins with a polarization intensity greater than 6$\sigma$, where some minority of data are in orthogonal polarization modes. The mean PAs are given in dots with error-bars of $\pm1\sigma$ of PA. The mean PA curve is fitted by the rotating vector model \citep{rc+1969} as indicated by the dashed line.}

\label{FigAverPol}
\end{figure}

\begin{figure}
\centering
\setlength\tabcolsep{0pt}
\begin{tabular}{ccc}
\includegraphics[width=0.45\textwidth]{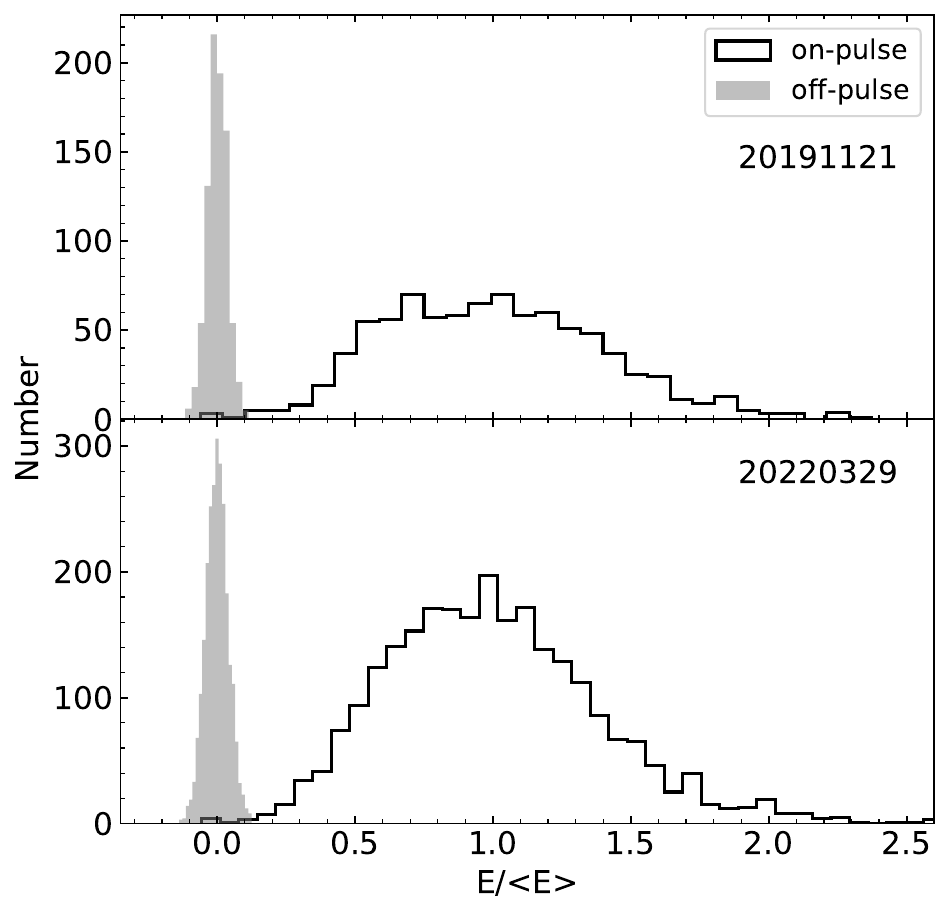}
\end{tabular}
\caption{The energy distribution of individual pulses of PSR J1857+0057 observed on November 21, 2019 (20191121, top panel) and March 29, 2022 (20220329, bottom panel). The pulse energy is counted in the longitude range in [$-25^{\circ}$, $20^{\circ}$] for pulse profile. For comparison, the energy distribution for the off-pulse range with the same size in [$-145^{\circ}$, $-100^{\circ}$] is plotted.
}
\label{Fighist}
\end{figure}

\section{Observations and data processing}
\label{sect:ObsDataPro}

PSR J1857+0057 was observed by the FAST on two sessions. One is done by chance for 5 minutes on  November 21, 2019 (hereafter 20191121)  during the FAST Galactic Plane Pulsar Snapshot (GPPS) survey \citep{Han2021}. The data of XX and YY channels are recorded with the time resolution of 49.152 $\mu s$ for 4096 channels covering the frequency range of 1000 to 1500 MHz. The interesting variable driftings of subpulses are found from the data of individual pulses (see Figure~\ref{pulse20191121}). The other session is the following-up observation carried out on March 29, 2022 (hereafter 20220329) for 15 minutes with the same time resolution but 2048 channels with polarization data recorded for XX, YY, Re[X*Y] and Im[X*Y], see \citet{Han2021} for details on setup.

Off-line data processing was carried out by using DSPSR \citep{van2011} and  PSRCHIVE \citep{Hotan2004} packages. The data are folded by DSPSR for single pulses with 512 phase bins per pulsar period. The  radio frequency interference (RFI) is excised by using PSRCHIVE. After these steps, the data are Faraday rotation corrected and dedispersed to a single pulse series, and data of both sessions have a high single-to-noise ratio to analyse single pulses.

\subsection{Polarization and geometry}

Following the procedures of \citet{Wang2023}, we get the mean polarization profile of PSR J1857+0057 and the position angle (PA) distribution for linear polarization detected for all bins of individual pulses in the second FAST observation session on 20220329, as shown in Fig~\ref{FigAverPol}. 
The single pulse sequences are presented in Fig~\ref{FigTimePhase1}, with full details of linear polarization percentage $L/I$, polarization position angle (PA) and the circular polarization percentage $V/I$ for all observation bins.
Compared with the mean polarization profiles in \citet{Weisberg1999}, our FAST data have a much higher signal-to-noise ratio due to the super sensitivity of the FAST. The weak emission around the longitude of +55$^{\circ}$ is also detected, in addition to the weak extended emission in the leading edge extended to about $-40^{\circ}$.

The PA curve of the mean pulse profile has the S-shape from the dominant polarization mode. The detailed PA distribution for all data bins, however, shows the second orthogonal polarization mode for a minority of data, in addition to the dominant data stream. In the leading edge in the longitude range of [$-25^{\circ}$, $-20^{\circ}$] the second orthogonal polarization mode becomes dominant. The mean PA curve between the longitude from -15$^{\circ}$ to 20$^{\circ}$ can be fitted by the rotating vector model \citep{rc+1969}. We get the fitted parameters: the magnetic inclination angle $\alpha_0 = 111.07^{\circ} \pm 5.17^{\circ}$, and impact angle $\beta_0 =  -9.15^{\circ} \pm 0.36^{\circ}$.

\subsection{No nulling detected}
\label{nonull}

The energy distribution of all detected individual pulses for PSR J1857+0057  observed on 20191121 and 20220329 are examined for nulling, as illustrated in Fig~\ref{Fighist}. Based on such super-sensitive observations, we find no indication for nulling during our two observation sessions.

\begin{figure}
\centering
\setlength\tabcolsep{0pt}
\begin{tabular}{ll}
\includegraphics[height=0.46\textwidth, angle=0]{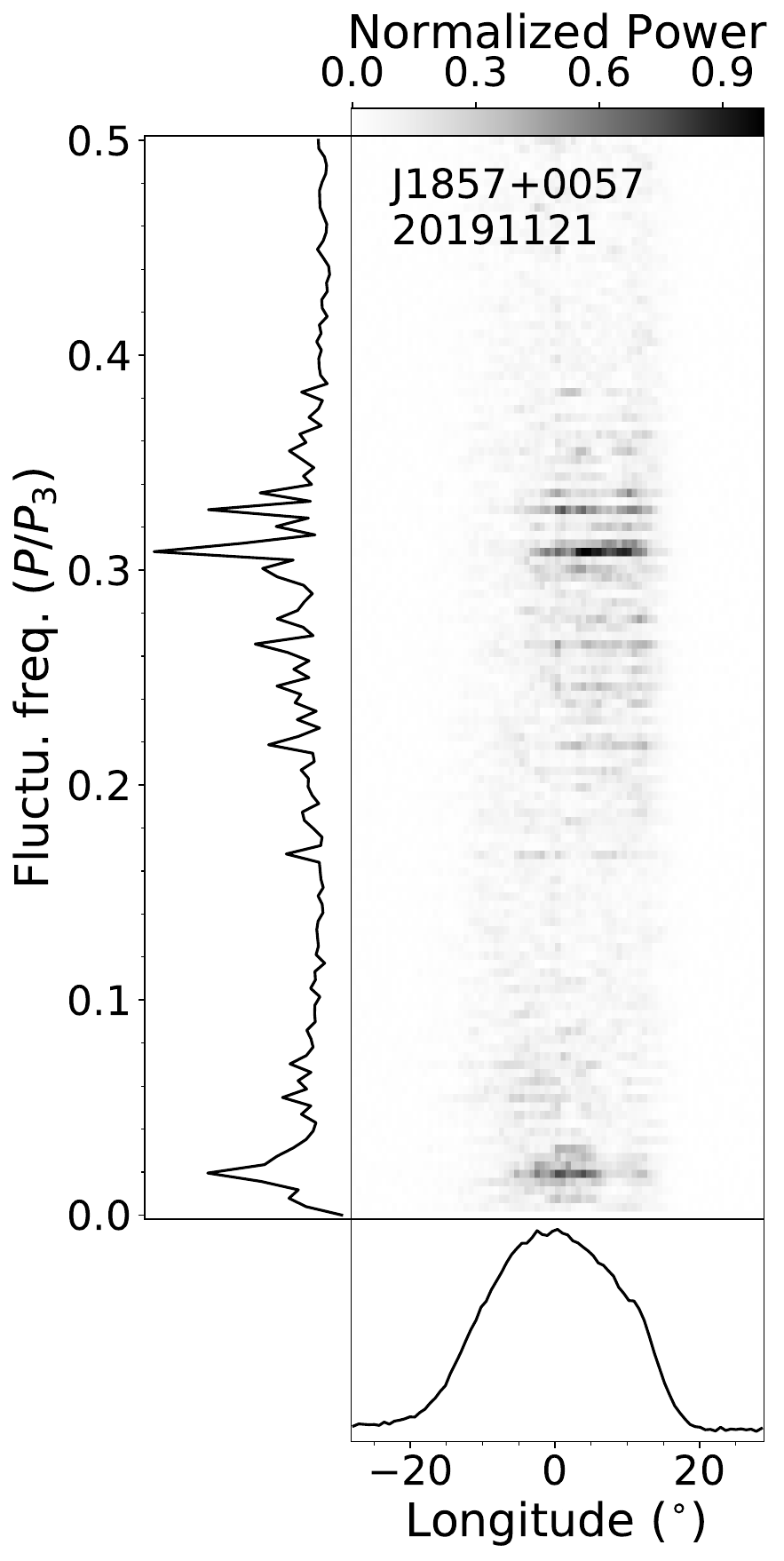}&
\includegraphics[height=0.46\textwidth, angle=0]{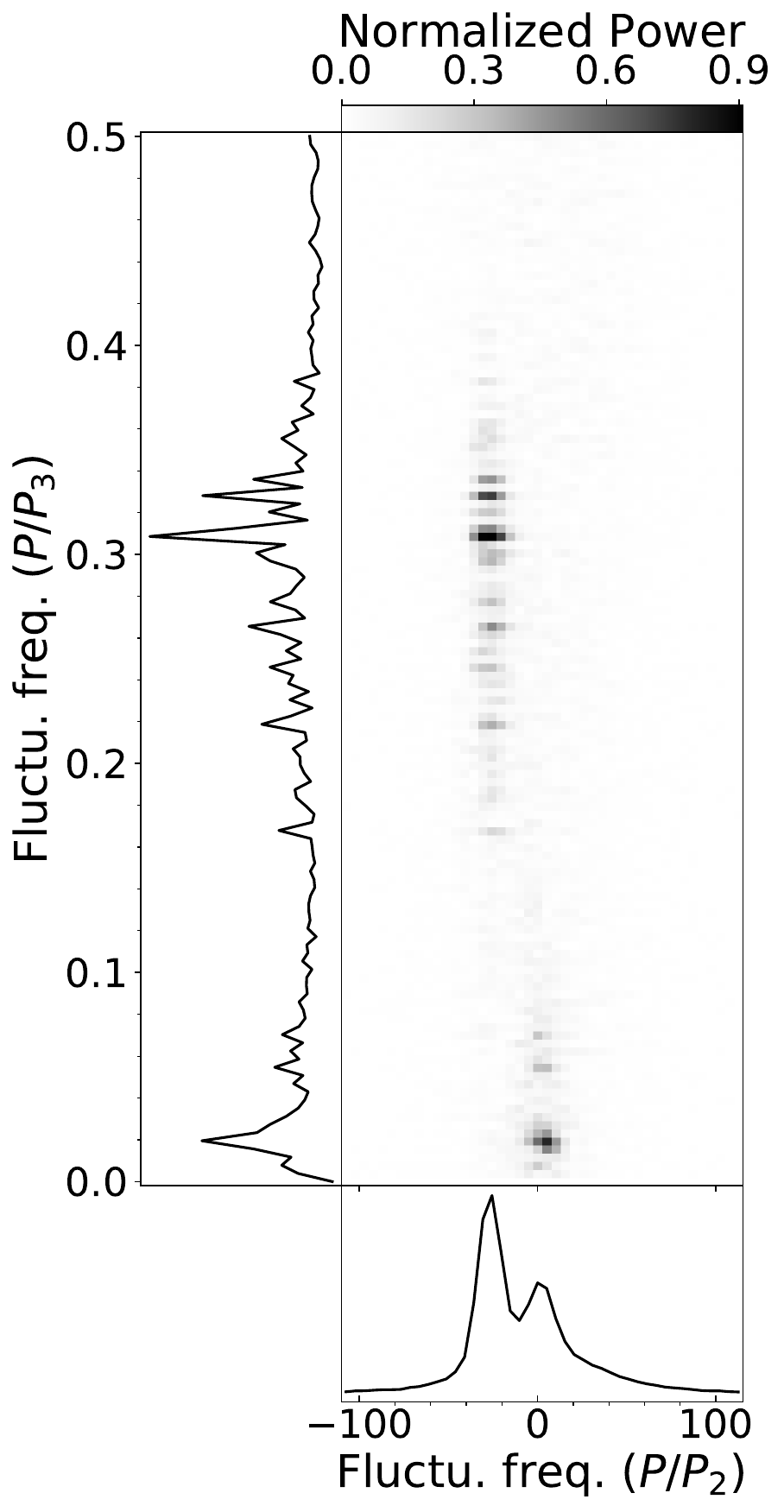}\\
\includegraphics[height=0.46\textwidth, angle=0]{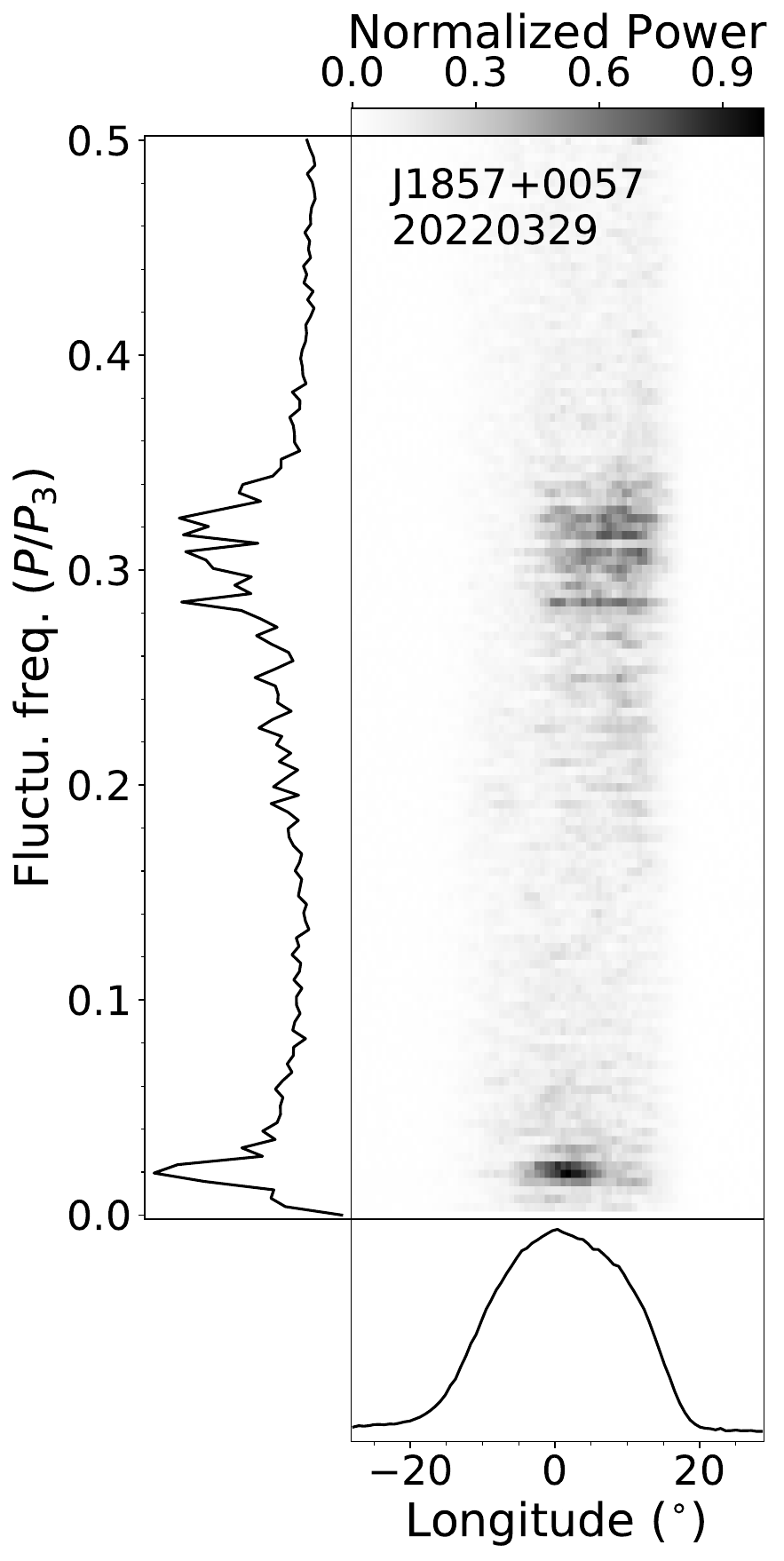}&
\includegraphics[height=0.46\textwidth, angle=0]{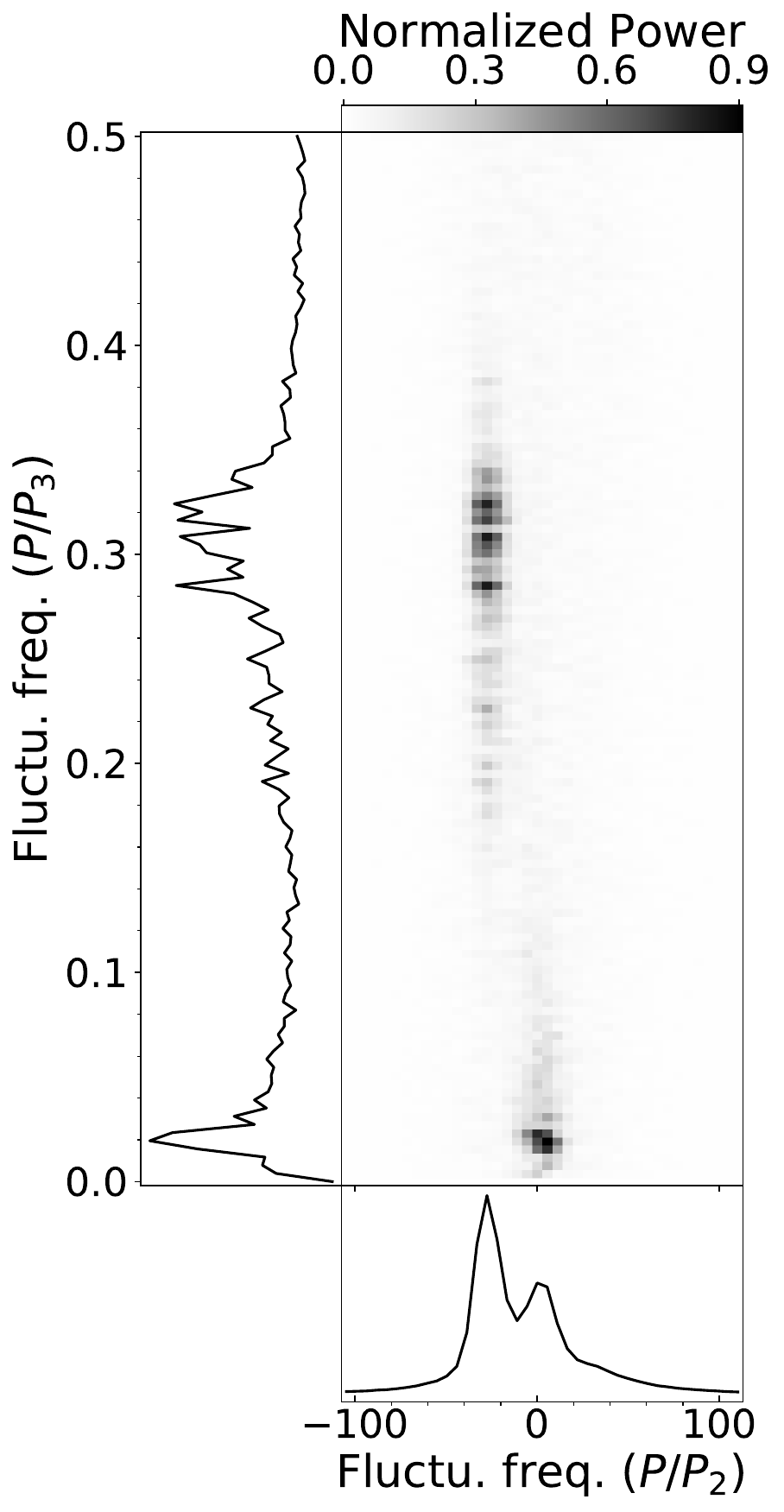}
\end{tabular}
\caption{Consistent results obtained from the LRFS ({\it left panels}) and 2DFS ({\it right panels}) analysis of single pulse series of PSR J1857+0057 for two observations on 20191121 ({\it top panels}) and 20220329 ({\it bottom panels}), both showing two groups of modulations, one for normal drifting at the fluctuation frequency of around 0.3 for $P_3\sim3.3P$, and the other is the low frequency modulation at the distinct fluctuation frequency of around 0.02 for $P_3'\sim50P$. }
\label{FigFS}
\end{figure}

\begin{figure*}
\centering
\setlength\tabcolsep{0pt}
\begin{tabular}{ccc}
\includegraphics[width=0.32\textwidth, angle=0]{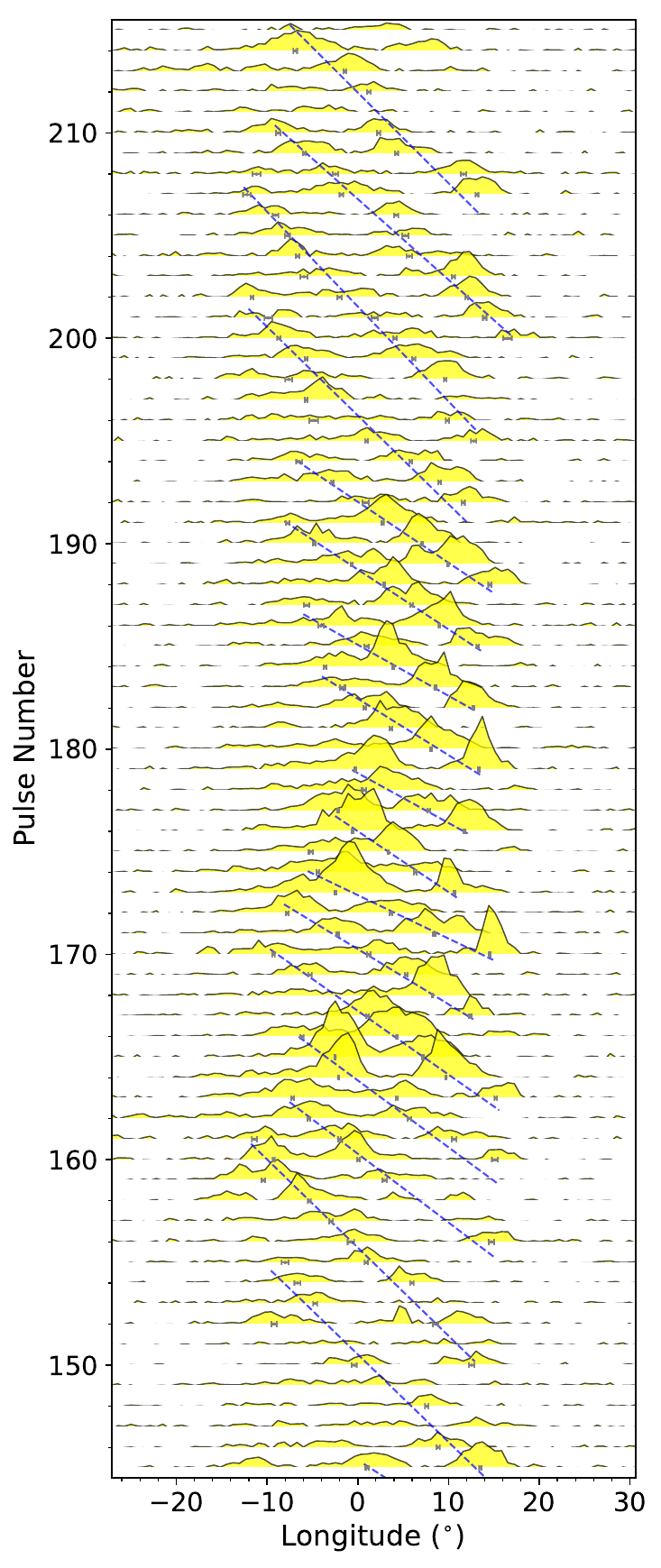}&
\includegraphics[width=0.32\textwidth, angle=0]{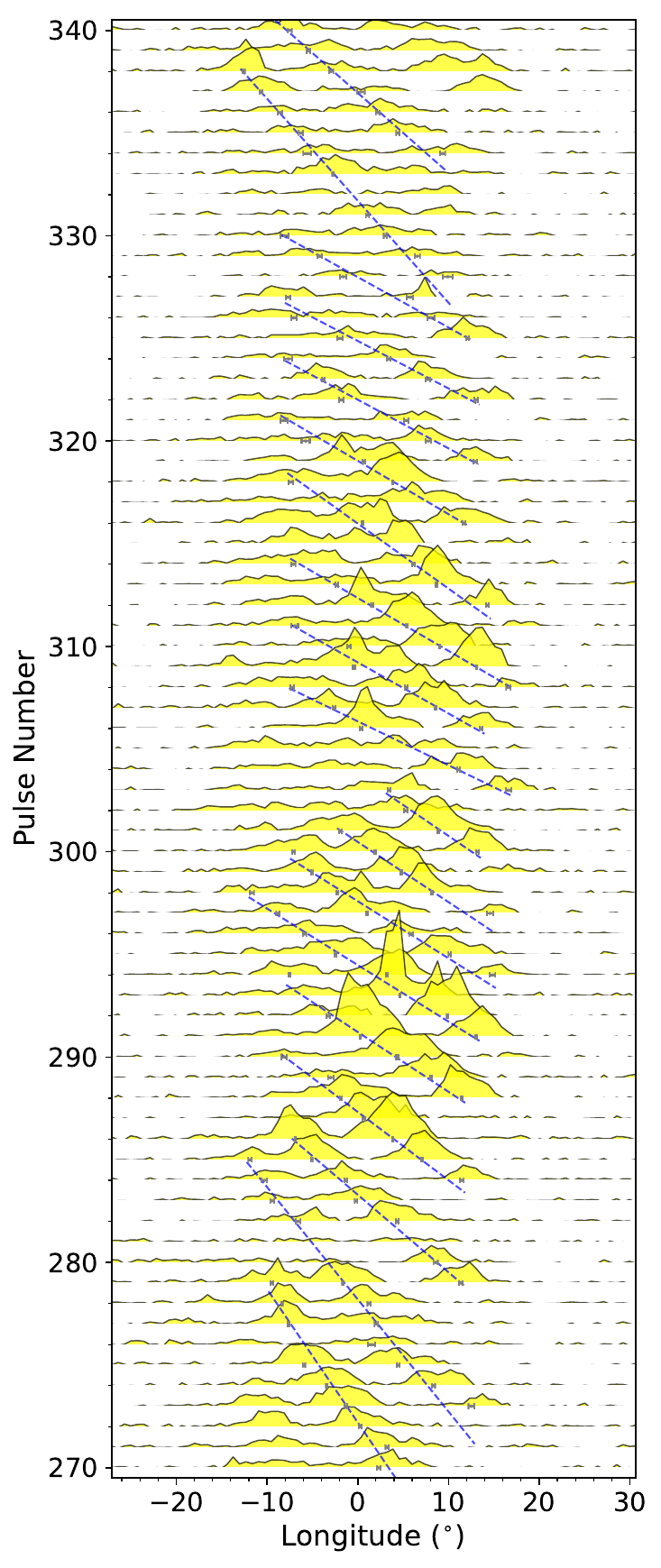}&
\includegraphics[width=0.328\textwidth, angle=0]{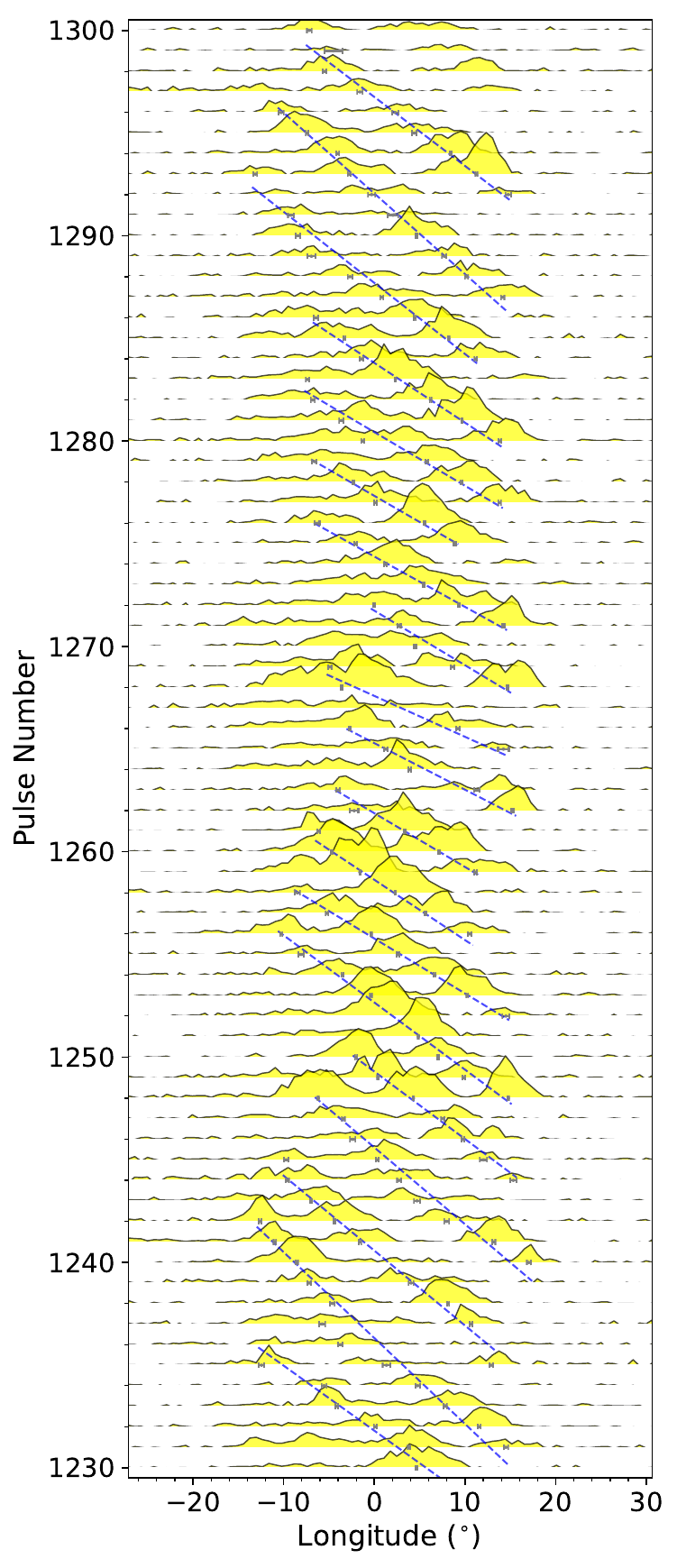}
\end{tabular}
\caption{Three segments of single pulse sequences observed in the session of 20220329, showing the similarly variable drifting of subpulses in a complete low frequency modulation cycle. Gray points represent the weighted mean phase of subpulses by pulse intensity with an uncertainty of $\pm 1 \sigma$, corresponding to locations of subpulses. Dashed lines are linear fitting to the mean longitudes of drifting subpulse of identified bands.}
\label{FigSinPulStacks}
\end{figure*}

\begin{figure*}
\centering
\setlength\tabcolsep{0pt}
\includegraphics[width=0.31\textwidth, angle=0]{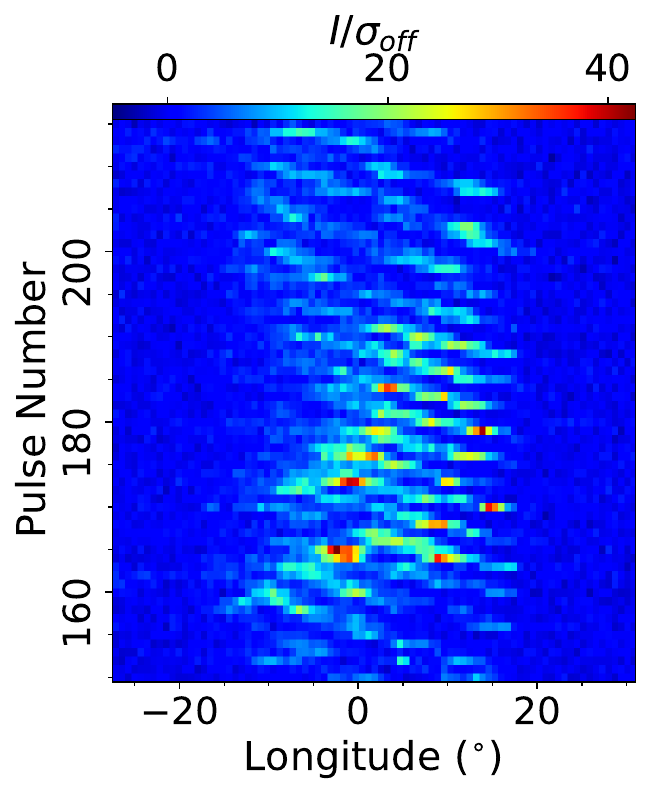}
\includegraphics[width=0.31\textwidth, angle=0]{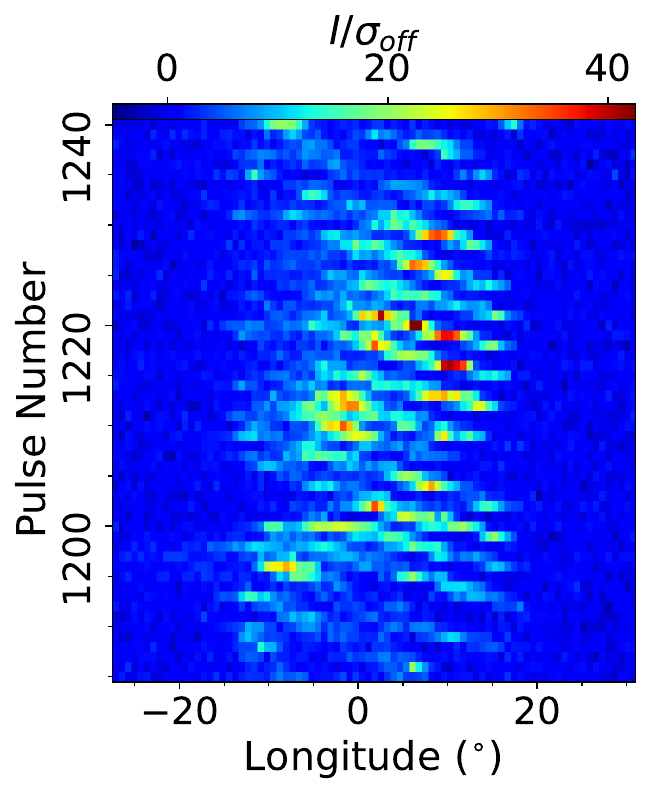}
\includegraphics[width=0.31\textwidth, angle=0]{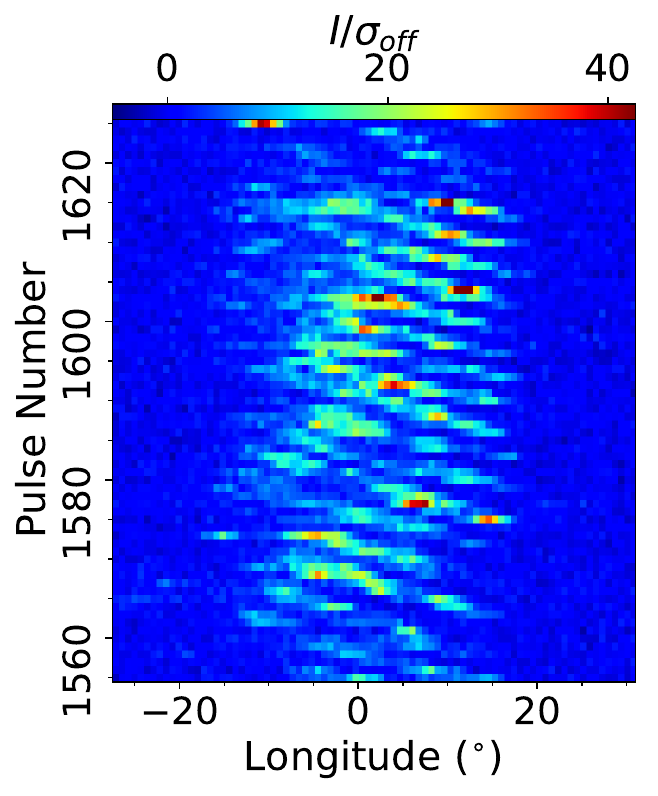}
\caption{Three segments of single pulse data observed in the session of  20220329, showing a global phase-forward intensity-enhancement (see $P_3'$ in Table~\ref{TabModuInfos}) in addition to the normal down-wards negative subpulse drifting (see $P_2'$ in Table~\ref{TabModuInfos}).}
\label{FigConfEll}
\end{figure*}

\begin{table}
\begin{center}
\caption[]{Fluctuation frequencies ($f_2$ and $f_3$) for subpulses of PSR J1857+0057 observed on 20191121 and 20220329 by FAST, both showing two groups of modulations. The peak and width values (and their uncertainties in brackets) of fitted Gaussian functions are listed. Drifting parameters $P_2$ and $P_3$ are derived from $f_2$ and $f_3$. }
\label{TabModuInfos}
 \begin{tabular}{lrrrr}
  \hline\noalign{\smallskip}
FAST Obs session &  \multicolumn{2}{c}{20191121} & \multicolumn{2}{c}{20220329} \\ \hline 
 Gaussian fitting    &  peak  &  width   & peak  &  width \\
\hline 
\multicolumn{5}{c}{subpulse drifting} \\
$f_3$ ($P/P_3$)    & 0.298(6) & 0.12(1) & 0.305(3) & 0.09(1)\\[1mm]
$f_2$ ($P/P_2$)    & -26.9(3) & 17.7(5)  & -27.4(3) & 18.4(7)  \\[1mm]
$P_3$ ($P$)        & 3.35(7)  &  0.25(3) & 3.28(3)  & 0.17(1) \\[1mm]                 
$P_2$ $(^{\circ})$ & -13.4(1) &  1.70(8)  & -13.1(1) & 1.70(8)    \\ 
\hline %
\multicolumn{5}{c}{Low frequency modulation} \\
$f_3'$ ($P/P_3'$)   & 0.020(1) & 0.016(2) & 0.021(1) & 0.019(2) \\[1mm]
$P_3'$ ($P$)       & 50(3)    &  8(2)   & 48(2)    & 9(1)    \\[1mm]          
\hline
\end{tabular}
\end{center}
\end{table}

\begin{figure*}
\centering
\includegraphics[width=0.9\textwidth, angle=0]{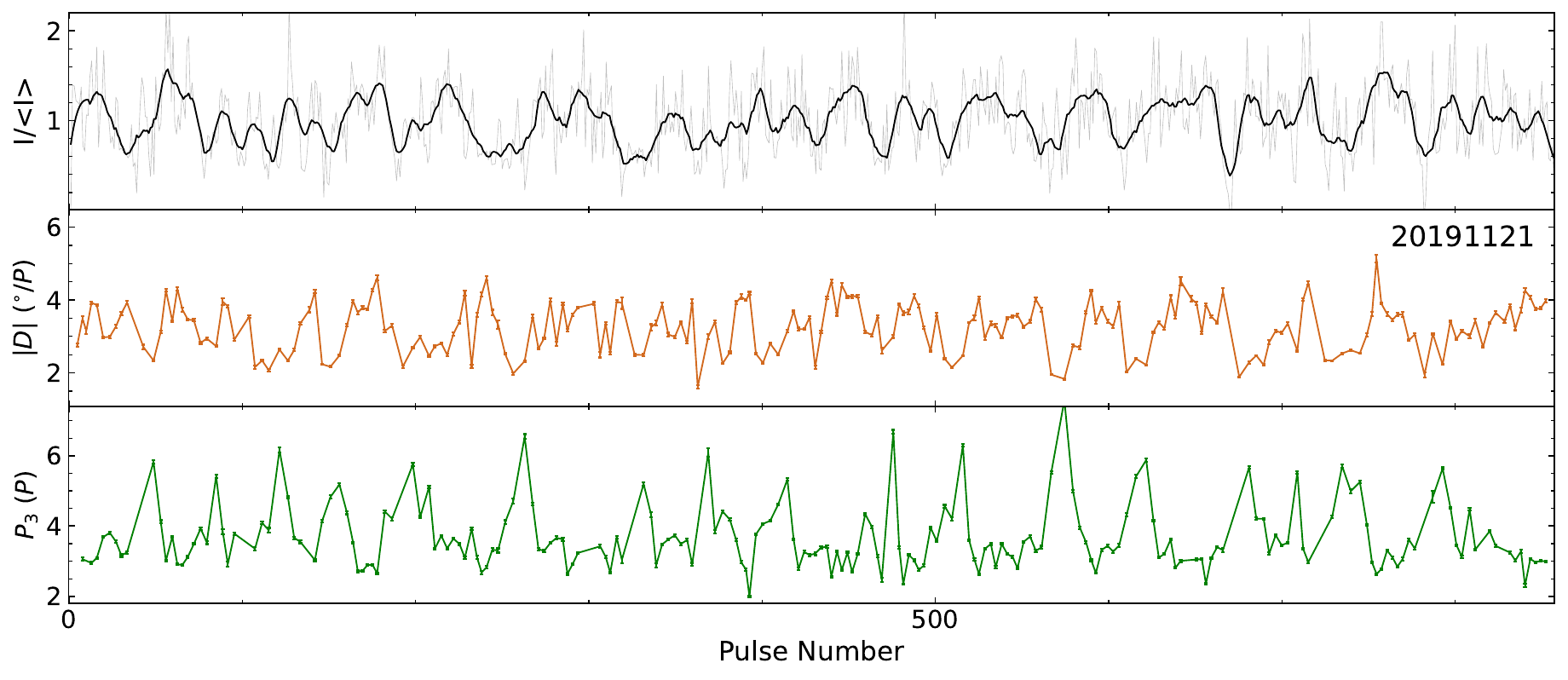}\\
\includegraphics[width=0.92\textwidth, angle=0]{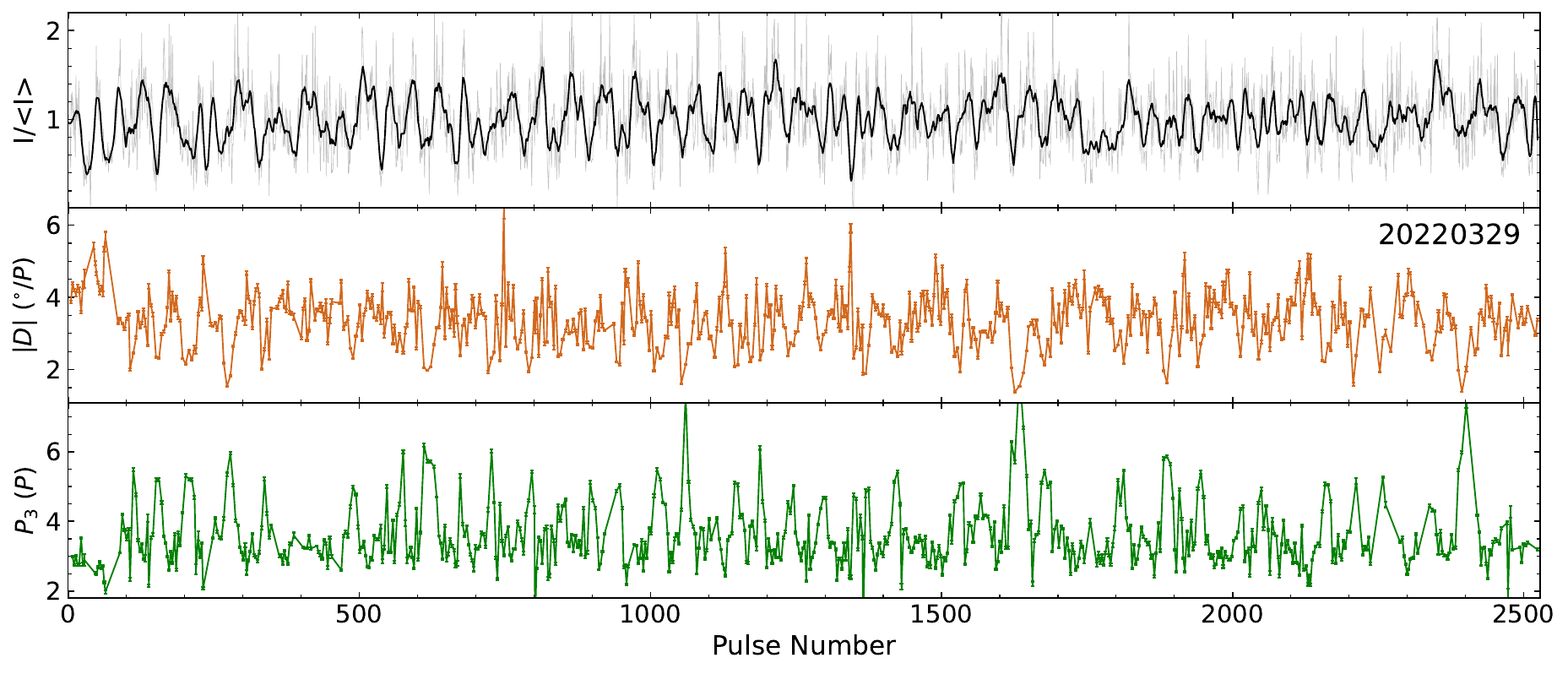} 
\caption{The pulse energy $I$ and smoothed line, the absolute drift rate $|D|$ and the time interval between drift bands $P_3$ with an uncertainty of $\pm 1 \sigma$ are plotted against pulse number for two FAST observation sessions on 20191121 and 20220329. }
\label{FigBandsInfos}
\end{figure*}

\begin{figure}
\centering
\includegraphics[width=0.325\linewidth, angle=0]{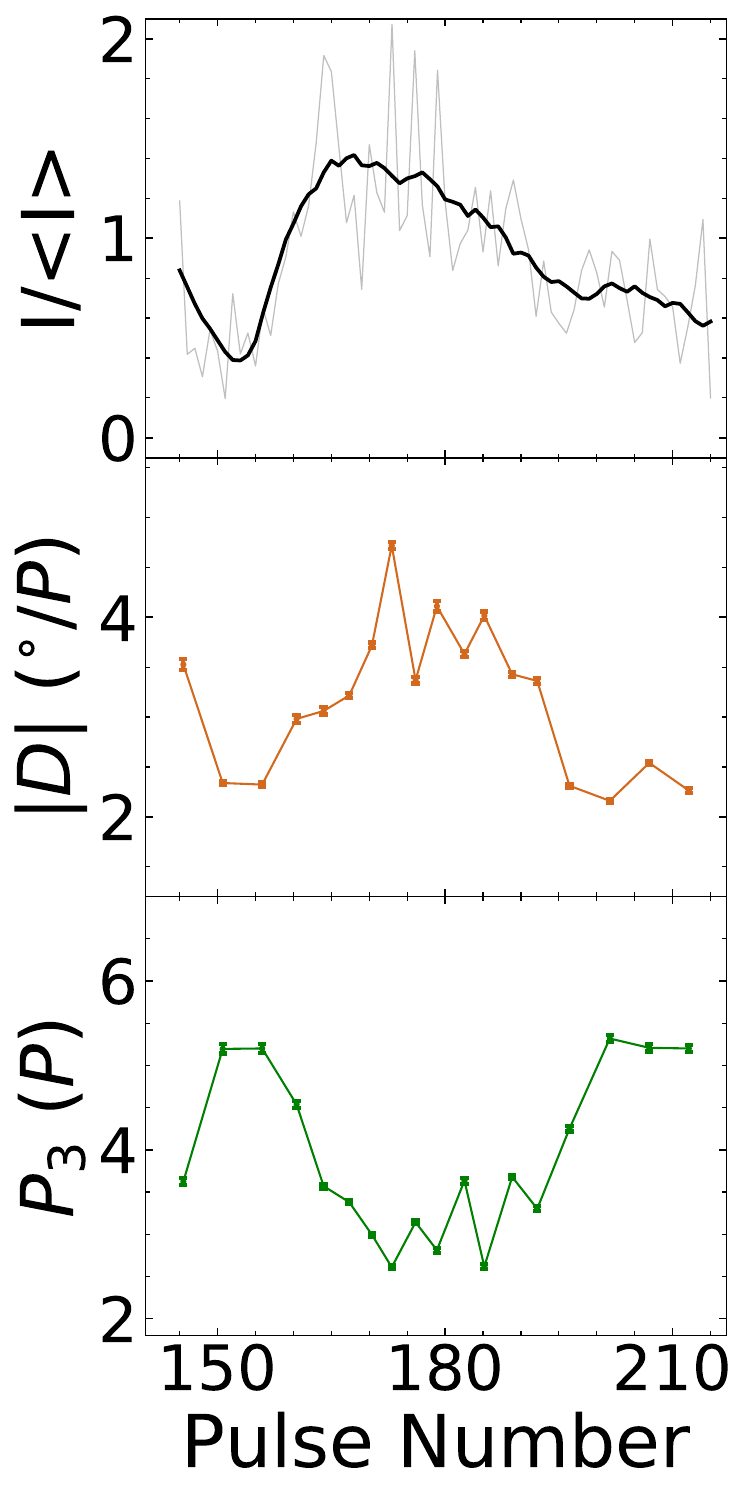} 
\includegraphics[width=0.325\linewidth, angle=0]{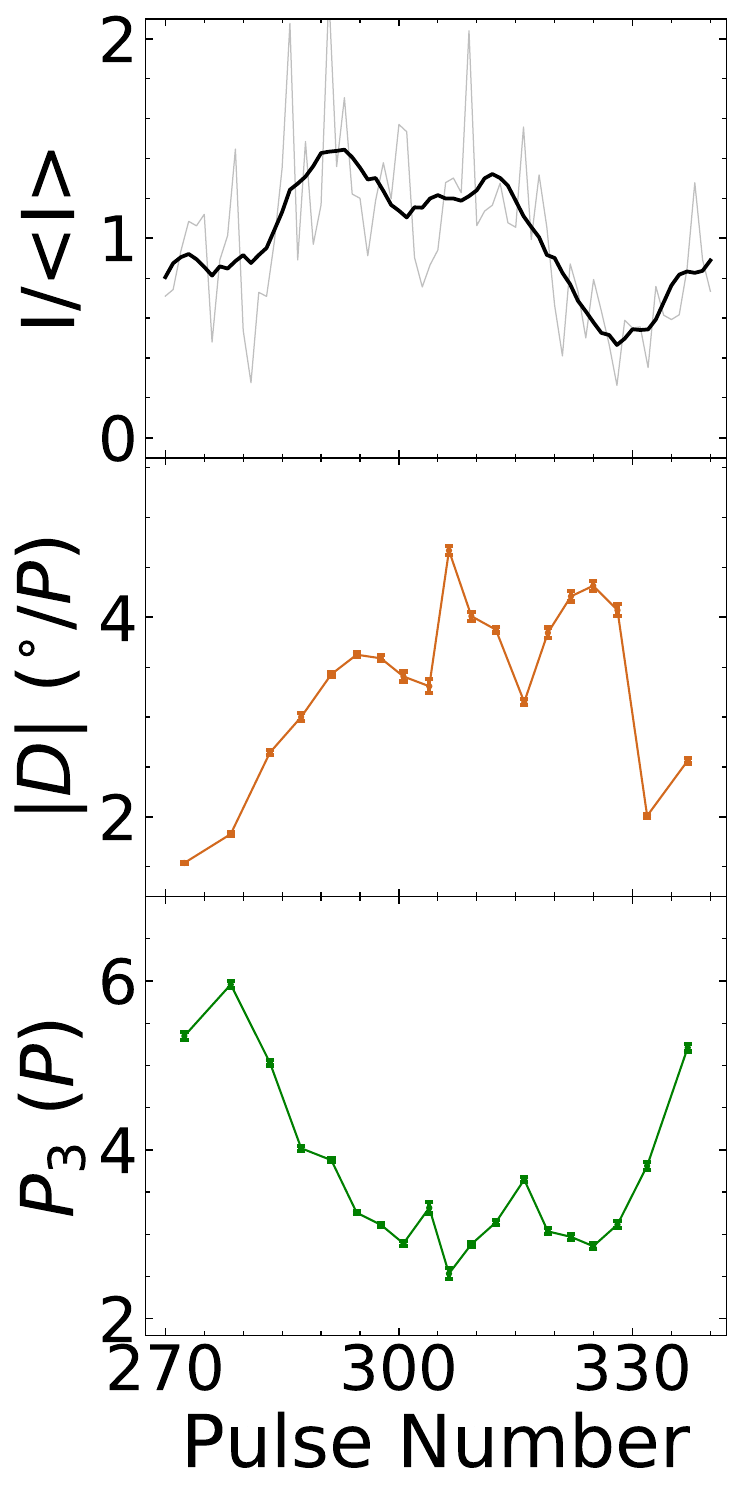} 
\includegraphics[width=0.325\linewidth, angle=0]{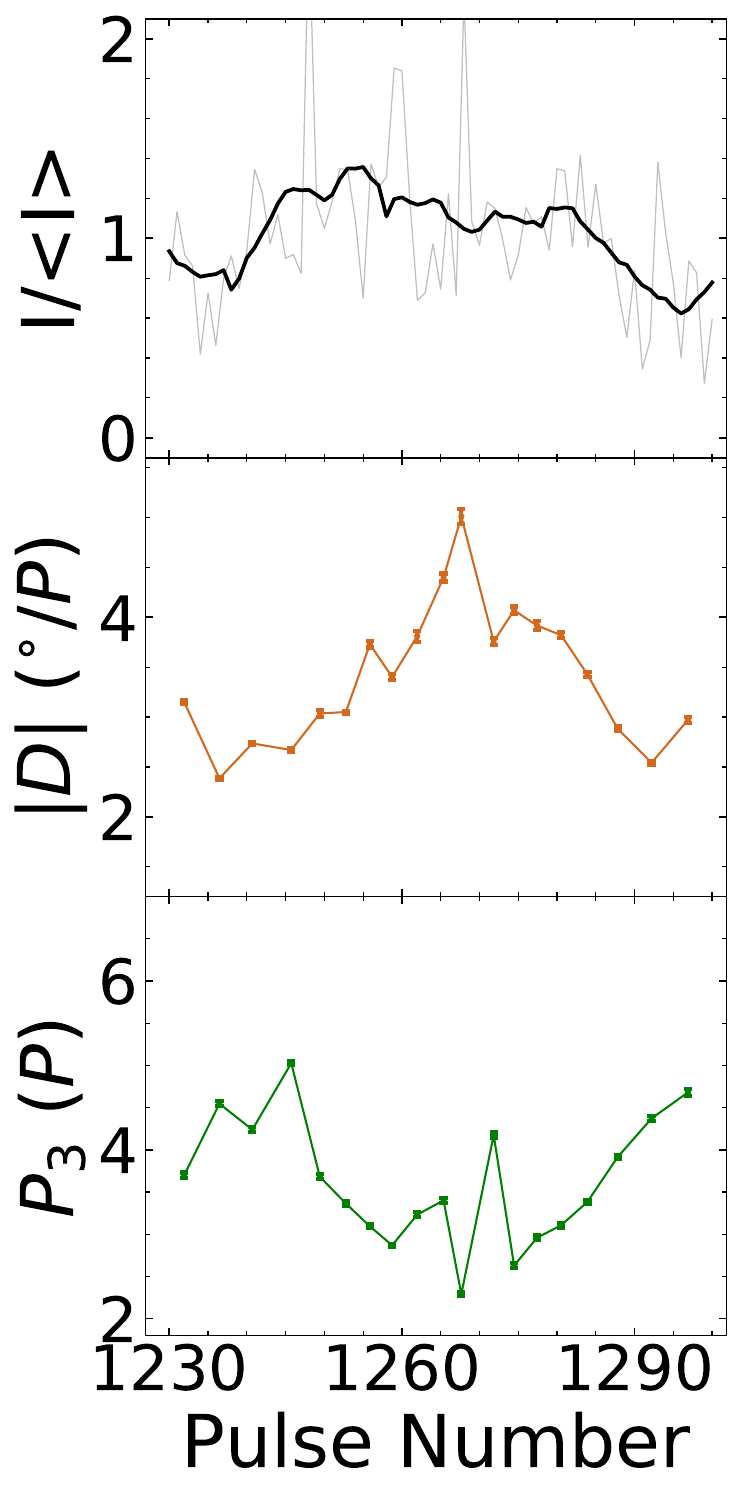}
\\
\caption{Same as Fig.~\ref{FigBandsInfos} but for three segments of data of pulses No.145-215, 270-340 and 1230-1300 in Fig~\ref{FigSinPulStacks}, showing clearly the parameter variations. }
\label{FigParsEvolu}
\end{figure}

\section{subpulse drifting and fluctuation analysis}
\label{sect:LowFreModu}

The methods for Longitude Resolved Fluctuation Spectra \citep[LRFS,][]{Backer1970} and the Two-Dimensional Fluctuation Spectra \citep[2DFS,][]{EdwSta2002} are used to analyse the drifting and modulation of subpulses. The LRFS is a discrete Fourier transform for data of every phase bin from the single pulse stacks, and can show the temporary modulation of each phase bin. The 2DFS is to do the two dimensional Fourier transform, and can get a global spectrum to assess the time and phase modulation frequencies. and the existence of drift feature. In practice, the package of PSRSALSA \citep{Weltevrede2016} is adopted to do both LRFS and 2DFS analysis.

Fig~\ref{FigFS} shows the results from the LRFS and 2DFS of PSR J1857+0057 for two observations on 20191121 and 20220329. The fluctuation features are consistent, as proved by similar drifting periodicities and fluctuation frequencies listed in Table~\ref{TabModuInfos} within the uncertainties. The integral fluctuation frequency curves in left and bottom panels of 2DFS in Fig~\ref{FigFS} can be smoothed and fitted by two-Gaussian functions. The fluctuation peak frequencies and the extension width are given in Table~\ref{TabModuInfos}. Here the width is the full-width at half maximum for the fitted Gaussian functions. Clearly, there are two groups of modulations. The first is the normal drifting mode, with a higher-frequency modulation of $P/P_3$ around 0.3, i.e. a periodicity of near 3.3$P$. It has a wide range of fluctuation frequency (see Fig.~\ref{FigSinPulStacks}), and the corresponding phase-modulation periodicity ($P_2$) is about $-13^{\circ}$. 
More interesting is the low frequency modulation, which has a fluctuation frequency about 0.02, i.e. the fluctuation periods of about every 50 periods. This low frequency modulation feature in 2DFS is asymmetric about the vertical axis, indicating that its phase modulation is opposite to the fast drifting direction and is related to global phase-forward intensity-enhancement which we will discuss in section~\ref{subsect:evolOfBands}.

Some individual pulses are shown in Fig~\ref{FigSinPulStacks} as examples for complete cycles of the low frequency modulation with variable drifting rate of subpulses, and the pulses for more periods in the color images in Fig.~\ref{FigConfEll} illustrate the global phase-forward intensity-enhancement.

\begin{figure}
\centering
\includegraphics[width=0.245\textwidth, angle=0]{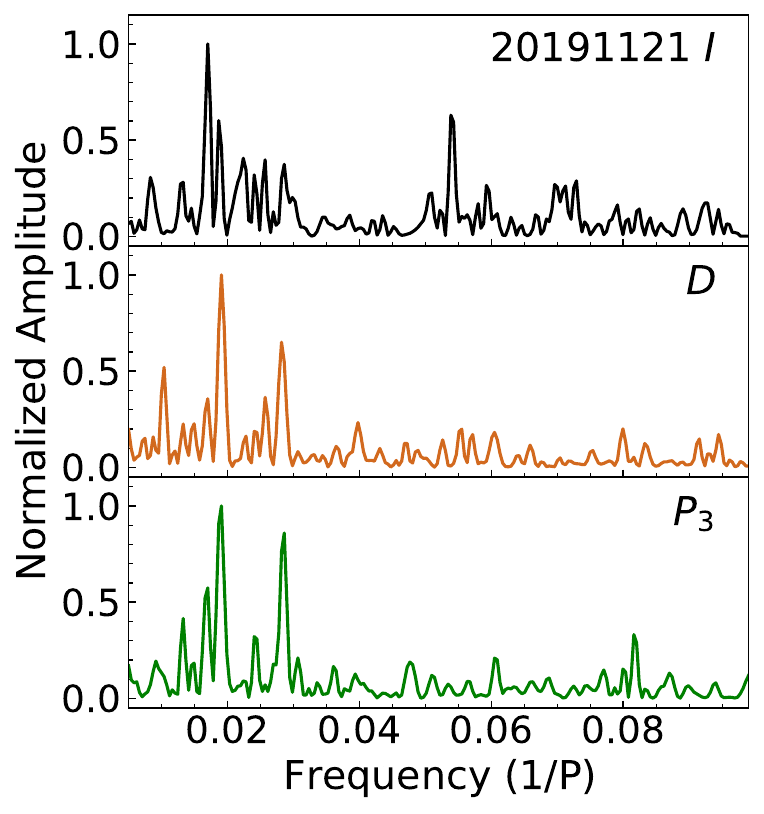}
\includegraphics[width=0.23\textwidth, angle=0]{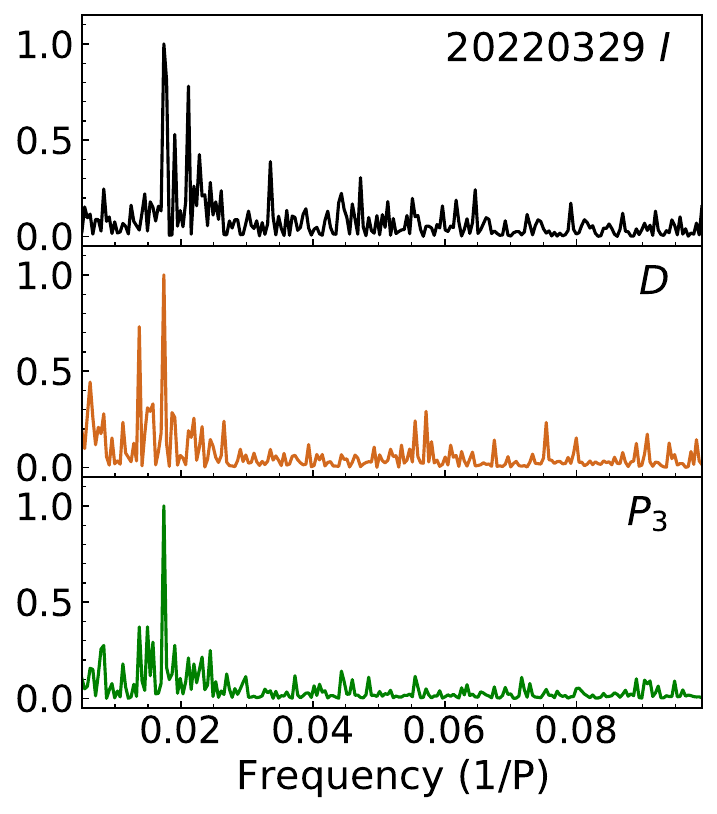}
\caption{The Lomb-Scargle periodograms for the variations of pulse energy I, drift rate $D$ and time interval $P_3$ of drift bands in Fig~\ref{FigBandsInfos}. The highest modulation frequency for $I$, $D$ and $P_3$ for 20191121 are found around 0.02, which correspond to the low frequency modulation of about 50 periods and have a false probability of 7.5e-09, 1.4e-03 and 3.4e-03 respectively, and that for 20220329 have a false probability of 3.6e-11, 3.4e-04 and 8.4e-07 respectively.}
\label{FigLSPeriodCrossCorre}
\end{figure}

\subsection{Variable drifting and low frequency modulations}
\label{subsect:variaDrift}

To reveal the variable drifting of subpulses, the central phase of a subpulse $P_c$ is estimated as the weighted mean of longitude $p_i$, using pulse intensity $I_i$ as weight of each bin of subpulse, 
\begin{equation}
P_c=\frac{\sum{(I_i p_i)}}{\sum{I_i}},
\end{equation}
so that the integrated energies of the subpulse on both sides are equal. 
See Appendix B for details. 

Based on the central phases of subpulses, we then identify drift bands, as shown in Figure~\ref{FigSinPulStacks}. The central longitudes of subpulses in a number of periods, if identified as a band of drifting subpulses, are linearly fitted, so that the slope is determined as the drifting parameter $1/D$. The uncertainties of the central phases of subpulses are used for the weights for this linear fitting. 
We assign the so-obtained $D$ to the given period among several periods for the fitted line very close to the longitude phase of $0^{\circ}$. 
The periodicity $P_{3}$ is obtained as the pulse numbers between the two fitted lines for two neighbored drifting bands at the pulse phase of $0^{\circ}$. Note that these two parameters are obtained  from the fitting of the individual pulses.
We also integrate the pulse energy $I$ in the longitude range of $-25^{\circ}$ to $20^{\circ}$ (see Fig.~\ref{FigAverPol}, scaled to the average) and then smoothed and plot the variations for two sessions in Fig~\ref{FigBandsInfos}, together with the absolute drift rate $|D|$ and the time interval between drift bands $P_3$ with an uncertainty of $\pm 1 \sigma$ against pulse number. Their variations for three specific segments are shown in Fig~\ref{FigParsEvolu}.

As seen in Fig~\ref{FigBandsInfos} and the detailed examples in Fig~\ref{FigParsEvolu}, quasi-regular variations of the drift rate $D$ and the temporary interval of adjacent drift bands $P_3$ are evident. 
Such periodic signals are obtained from the Lomb-Scargle periodogram for the fluctuations of pulse intensity $I$, drift rate $D$ and time interval $P_3$ of drift bands. The modulation periodicities indicated by the peaks are around 50$P$. The false probabilities of these peaks are less than 0.5\% \citep{Baluev2008}. Note that the low frequency modulation are not only shown in pulse intensity, but also from the drifting parameters.
Because PSR J1857+0057 has no nulling (see Sect.~\ref{nonull}),
the low frequency feature could be caused by intensity modulation, as very clearly shown in a set of color plots in Fig.\ref{FigTimePhase1} in the appendix.

\subsection{Within a low frequency modulation cycle}
\label{subsect:evolOfBands}

To show the drifting variations within a low frequency modulation cycle, we take three segments as examples in Fig~\ref{FigParsEvolu} which show the variations of the integral intensity $I$, absolute drift rate $|D|$ and temporary interval $P_3$ in a complete low frequency modulation cycle. For most cycles, the integrated pulse energy $I$ and the absolute drift rate $|D|$ increase first and then decrease, while $P_3$ has a trend to decrease first and then back to larger values within a modulation cycle. 

\begin{figure}
\centering
\includegraphics[width=0.99\linewidth, angle=0]{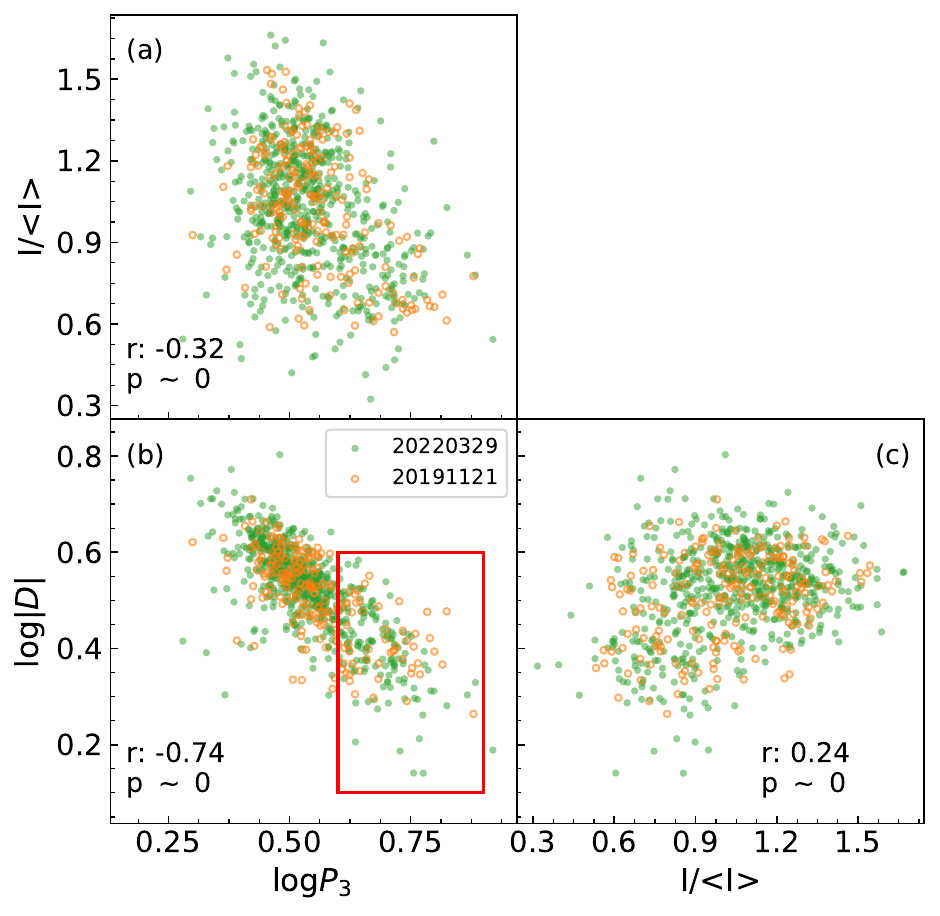}
\caption{The pulse-energy $I$,  log$|D|$ and log$P_3$ for observations on 20191121 (orange circle) and 20220329 (green point), showing a strong correlation between $D$ and $P_3$. Red box indicates the special ranges for $P_3$ and $D$ in Fig~\ref{FigPAContrast}.}
\label{FigCorrIShiftDrift}
\end{figure}

\begin{figure}
\centering
\includegraphics[width=0.35\textwidth, angle=0]{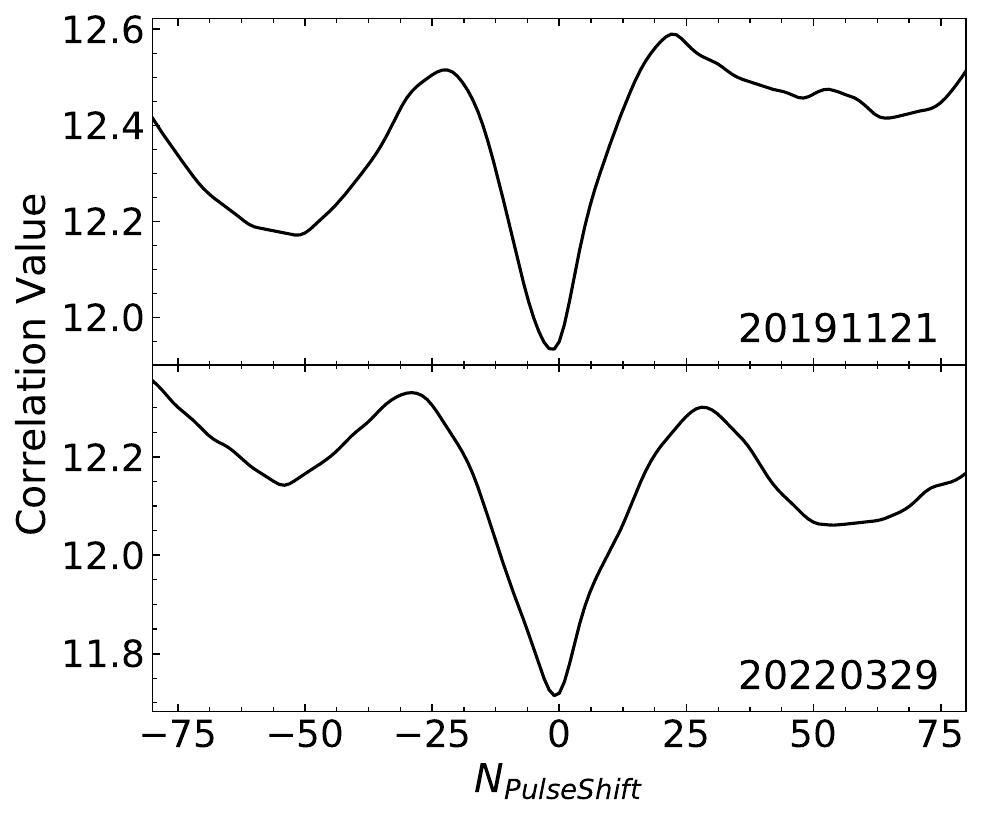}
\caption{The anti-correlation of data series of $|D|$ and $P_3$ for variations in Fig.~\ref{FigBandsInfos} is clearly detected at the shifted-pulse number $N_{\rm PulseShift}\simeq0$.
}
\label{FigCrossCorre}
\end{figure}

\begin{figure}
\centering
\includegraphics[width=0.93\linewidth, angle=0]{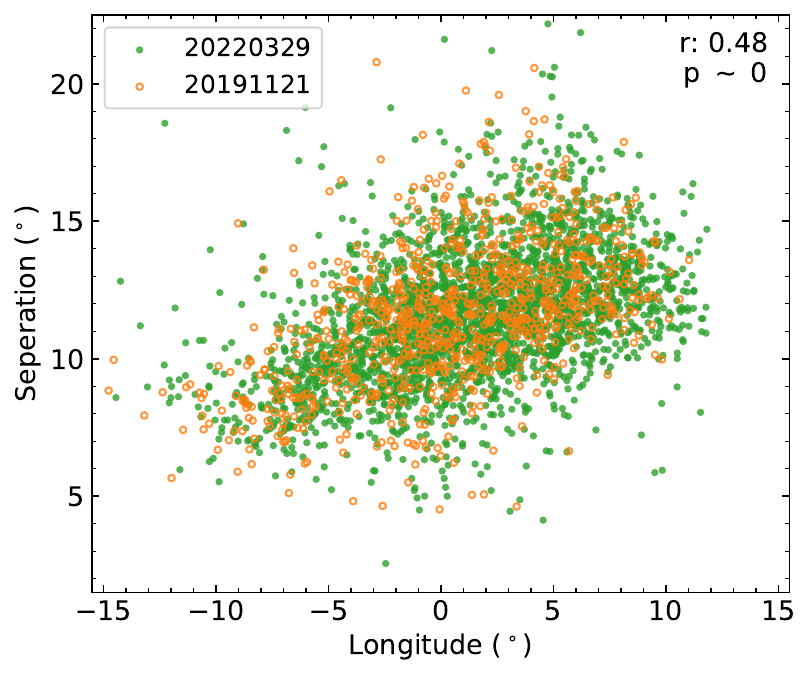}
\caption{The phase separation of two adjacent subpulses versus the center longitude from observations on 20191121 (orange circle) and 20220329 (green point).}
\label{FigCorrP2Phase}
\end{figure}

Within a modulation cycle, a global phase-forward intensity enhancement feature is shown in Fig~\ref{FigConfEll}, adding a clear modification of the subpulse intensity in a normal drifting band. Strong subpulses appear in a later longitude phase in the next band. As a consequence, the apparent phase-forward intensity-enhancement feature is often visible as a global pattern though it is often not stable. 

Note, however, that the low frequency modulation are not stable, only quasi-regular. First, such changes of drifting parameters shown in  Fig~\ref{FigParsEvolu} are presented for many cycles, but not the same for each cycle. Secondly, the phase-forward intensity-enhancement appears in some cycles, not in all cycles. 


\subsection{Correlations between parameters for subpulse drifting}

If the drifting is not variable, i.e. the $D$ is fixed to one value, it would be easy to measure $P_2$ and also $P_3$, and obtain $D = P_2 / P_3$. However, this pulsar has variable drifting rates $D$, obtained directly from fitting the drifting bands, so that one can see a wide distribution of $D$ in Fig.~\ref{FigCorrIShiftDrift}. On the other hand, the values of $P_2$ are diverse (even for two values in a given period), though the averaged values are given in Table~\ref{TabModuInfos}. These drifting parameters are no longer confined by the simple relation $D = P_2 / P_3$. Nevertheless, since $P_2$ has a mean value, the negative correlation between $|D|$ and $P_3$ is expected.

By looking at the variations of the absolute drift rate $|D|$ and the temporary separation  $P_3$ between adjacent drift bands, one can see that they are anti-correlated, and the cross-correlation is shown in Fig.~\ref{FigCrossCorre} confirming the anti-correlation with the shifted pulse number $N_{\rm PulseShift}\simeq0$.

To further examine the possible numerical dependence between the on-pulse integral intensity $I$, drift rate $D$ and temporary interval $P_3$, we plot all data in Fig~\ref{FigCorrIShiftDrift}. 
The Spearman correlation coefficients $r$ and associated p-values $p$ are calculated, with null hypothesis that two quantities are uncorrelated. The p-values are all very close to 0 indicating significant correlation. We get the correlation coefficients of -0.32, -0.74 and 0.24 respectively in panels (a), (b) and (c), and conclude the strong negative numerical correlation with a slope of around $-1$ between log$|D|$ and log$P_3$. The data scattering is not a reflection of the uncertainty in determination of subpulse locations, but due to diverse $P_2$, which has the tendency to be larger in the later longitudes as shown Figure~\ref{FigCorrP2Phase}. Data are more scattered in panels with intensity $I$, indicating weaker correlations.

\begin{figure}
\centering
\setlength\tabcolsep{0pt}
\begin{tabular}{ccc}
\includegraphics[width=0.45\textwidth, angle=0]{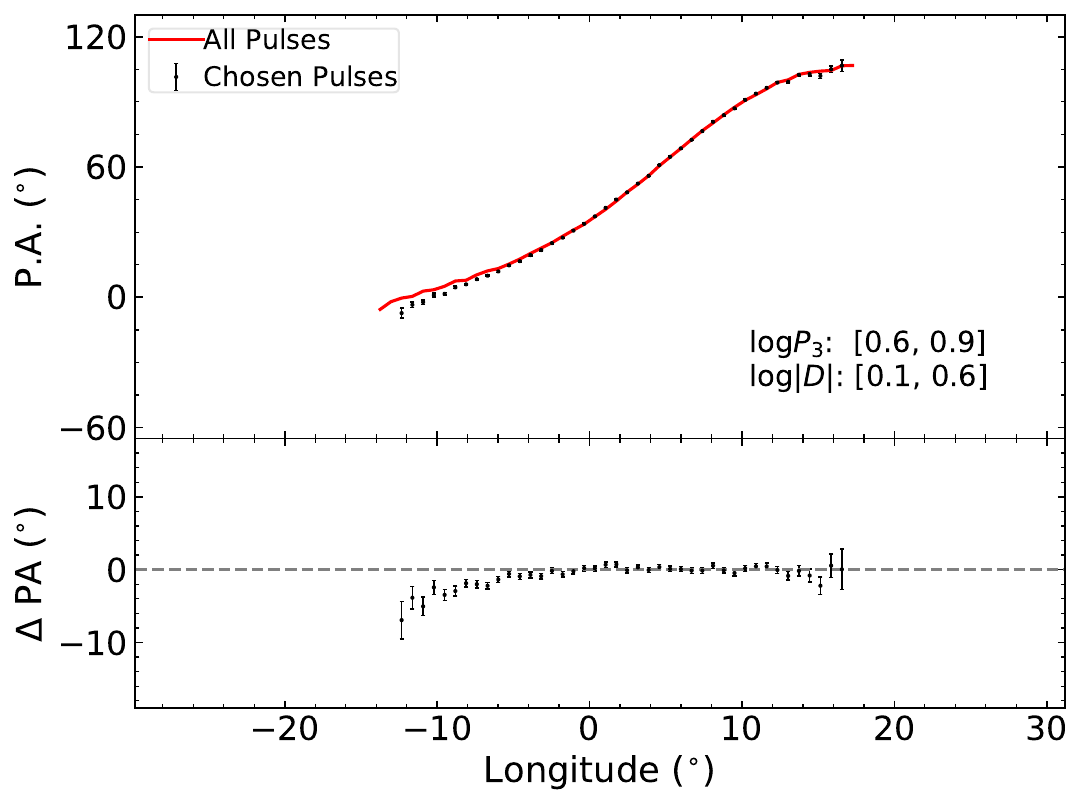}
\end{tabular}
\caption{Unusual polarization angles for less drifting pulses, with a large log$P_3$ in the range between 0.6 and 0.9 and a small log$|D|$ in the range between 0.1 and 0.6 in the marked red box in Fig~\ref{FigCorrIShiftDrift}. They have steep PA curves at the leading edge of the mean profile, compared to the mean of all pulses.}
\label{FigPAContrast}
\end{figure}




\subsection{Unusual polarization for less drifting pulses}

In each cycle, drift bands start with a smaller drift rate $|D|$ and larger time interval $P_3$. We checked the polarization data for these individual pulses, and found that single pulses with smaller drift rates (log$|D|$ from 0.1 to 0.6) and larger time intervals (log$P_3$ from 0.6 to 0.9) have  smaller PA values in the leading edge of the mean profile (see  Fig~\ref{FigPAContrast}). Comparing with the average PA curve for all pulses, the PA of these less-drifting pulses have a difference of several degrees. 

\section{Discussion and conclusions}
\label{sect:Discussion}

We have shown for the first time that PSR J1857+0057 has a low frequency modulation of about $50P$, in addition to the normal negative drifting with $P_3\sim3.3P$. From the Lomb-Scargle periodograms, we found that the low frequency modulation is not only shown in intensity $I$, but also on drift parameters. Obvious cross-correlations are found between $D$ and $P_3$, mainly due to the low frequency modulation. In the modulation cycle, 
%
the integral intensity $I$ and absolute drift rate $|D|$ seem to increase first and then decrease, and the temporary interval of drift bands $P_3$ has the opposite tendency. 
Some cycles show apparent phase-forward intensity-enhancement from single pulse sequences that is consistent with the low frequency positive phase modulation feature in 2DFS.
%
%
%
However, the low frequency modulation are not stable. 
The phase-forward intensity-enhancement feature in time-phase graphs and evolution of intensity and drift parameters are sometimes, but not always, presented in a modulation cycle.

For quasi-periodic modulation in time-phase graphs, there are possibly two kinds of origins. One is the geometrically periodic change caused by some physical mechanisms, leading to the periodicity in observations. For instance, subbeams circulate around the magnetic axis leading to modulation on subpulses in time and phase periodically. The rotation of binary stars may also cause periodic phase modulation in time-phase graphs. The other origin is not related to geometry, but to self-regulating mechanism along with the periodic variations of physical quantities. 
In general, the periodic modulation related to geometry is more stable compared with the self-regulating mechanism.

\subsection{The carousel circulation modulation}


The low frequency modulation feature or some sidebands around the primary  feature in the fluctuation spectra are often interpreted as the tertiary modulation \citep{Deshpande2001,Mitra2008,Maan2014}, resulted from the rotating subbeam carousel cycle that is related to the carousel circulation period. If the characteristics of subbeams are distinct and stable, there will be the tertiary modulation based on the rotating carousel model. Otherwise, the features corresponding to the rotating carousel in the fluctuation spectra will be invisible such as for PSR B0809+74  \citep{Leeuwen2003,Rankin2006}.

As reported in Section~\ref{sect:LowFreModu}, the low frequency modulation of PSR J1857+0057 is not only shown for the intensity but also for drifting parameters including the drift rate. Though the rotating carousel model can be related to the low frequency modulation on drifting in terms of subpulse-properties evolution along drift bands, it is difficult to explain the periodic variations of drift parameters. We therefore suspect that the low frequency modulation of PSR J1857+0057 is not caused by the rotating carousel cycle.

\subsection{Drift rate variations}

The gradual changes of the drift rate have been previously reported for few pulsars \citep{Gupta2004,Esamdin2005,Leeuwen2012,Bhattacharyya2010,McSweeney2017,Lu2019}. Some studies explained the drift rate with the Partially Screened Gap (PSG) model \citep{Gil2003, Bhattacharyya2010, Gupta2004, McSweeney2017}. In the thermally self-regulated PSG model, the flow of iron ions from the polar cap surface partially screen gap. The stellar surface temperature influences the ions flow and further the electric field, which will lead to the variations of drift rate.
%
%
\citet{Leeuwen2012} presented the dependence of drift rate on the variation of the accelerating potential across the polar cap, and that the drift rate change of PSR B0826$-$34 is caused by the variation of accelerating potential in both space and time.  The diffuse drifting of PSR B2016+28 may be resulted from the rough stellar surface or the asymmetric shape of polar gap \citep{Lu2019}. 

However, such models can not fully explain the periodic and gradual change of drift parameters and apparent low frequency phase-forward intensity-enhancement for PSR J1857+0057.
%
%
Considering the instability of low frequency modulation, we suggest that the low frequency of PSR J1857+0057 is caused by the self-regulating mechanism.

\subsection{Correlations between the drifting parameters}

Some pulsars with variable drift rates were found to have correlations among different quantities. For example, \citet{Bhattacharyya2010} found that the intensity of PSR B0818$-$41 decreases gradually and the apparent drift rate slows down before the nulls. They suggested that the electric field in the polar gap is composed of parallel and perpendicular components to the surface magnetic field, which are respectively related to the intensity and drifting, and the variation of electric field will influence both of them. As a result, the correlation can exist between the intensity and drift rate. 

Intensity and drift parameters of PSR J1857+0057 show the modulation of around $50P$ in the periodograms. As shown in section~\ref{subsect:evolOfBands}, the intensity $I$ and absolute drift rate $|D|$ of PSR J1857+0057 both have the tendency of increasing first and then decreasing during low frequency modulation cycle, which may indicate that they are both caused by the change of electric field, as discussed by \citet{Bhattacharyya2010}.



\section*{Acknowledgements}


%
The authors are supported by the National Natural Science Foundation of China (NSFC, Nos. 11988101, 11833009 and 12133004). 

\section*{Data availability}
All data in this paper are available with kind request from authors.



\bibliographystyle{mnras}
\bibliography{ref.bib} 




\appendix

\section{Single pulse stacks of PSR J1857+0057}


Here we present the single pulse stacks of PSR J1857+0057 observed on 20191121 and 20220329 in Fig.~\ref{pulse20191121} and Fig.~\ref{FigTimePhase1}, together with their variations of the pulse energy $I$, drift rate $D$, the drifting periodicity $P_3$. The linear polarization percentage, polarization angle, and circular polarization percentage of observation on 20220329 are also shown in details. For clarity, each stack shows about 440 pulses.


\begin{figure*}
\centering
\begin{tabular}{ll}
\includegraphics[height=0.95\textheight, angle=0]{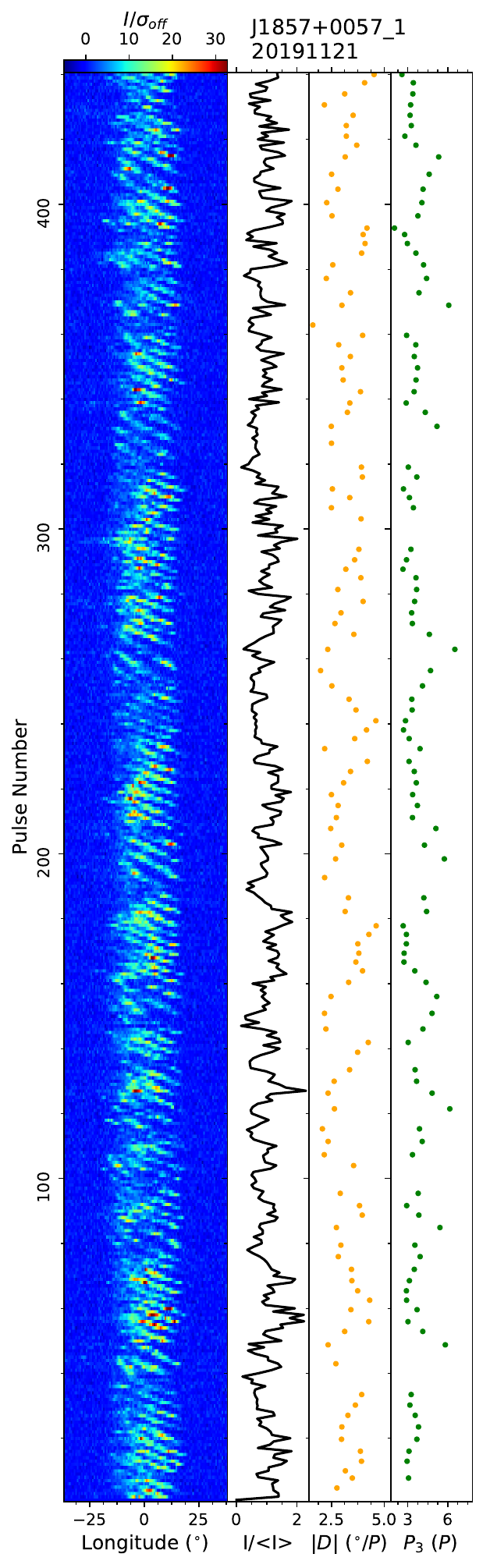}&
\includegraphics[height=0.95\textheight, angle=0]{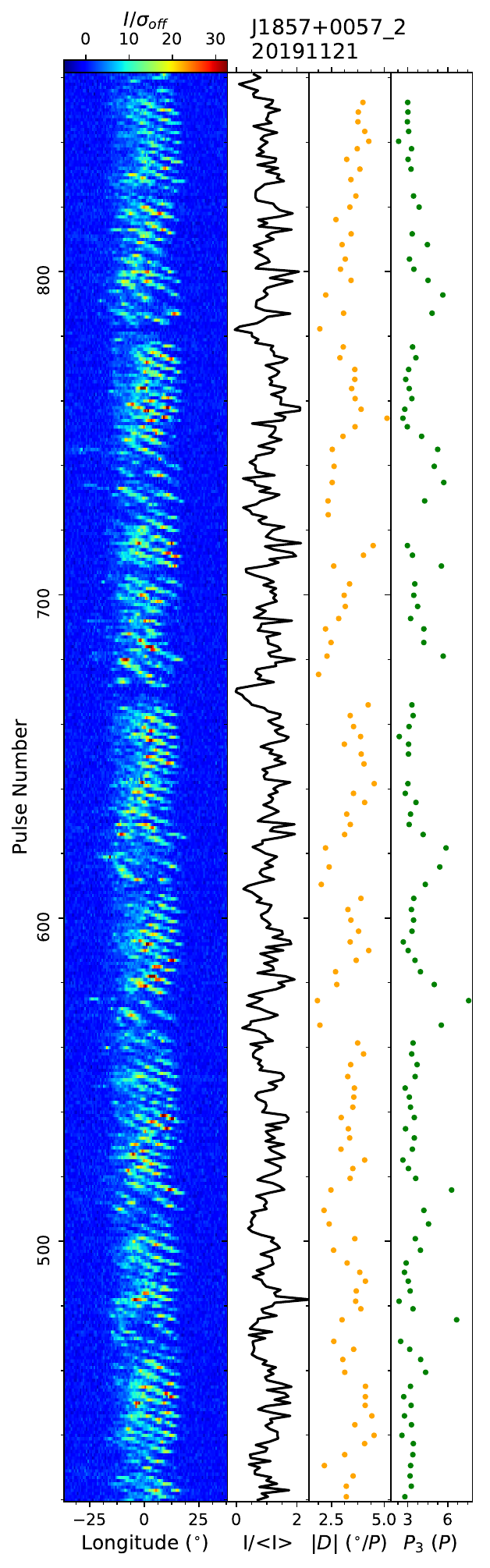}
\end{tabular}
\caption{Single pulse stacks of PSR J1857+0057 observed on 20191121 by the FAST, together with variations of the intensity of individual pulses $I$, drifting rate $D$, the drifting periodicity $P_3$. Quasi-regular variations are remarkable and quantitatively analysed in Sect.3. }
\label{pulse20191121}
\end{figure*}

\begin{figure*}
\centering
\includegraphics[height=0.95\textheight, angle=0]{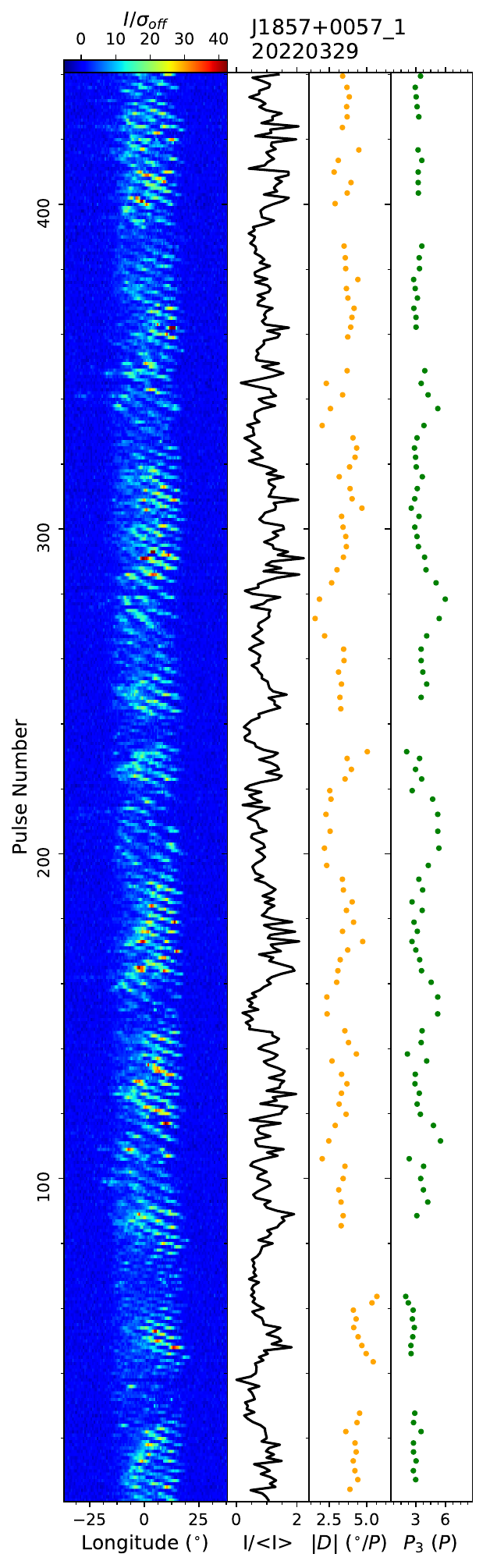}
\includegraphics[height=0.95\textheight, angle=0]{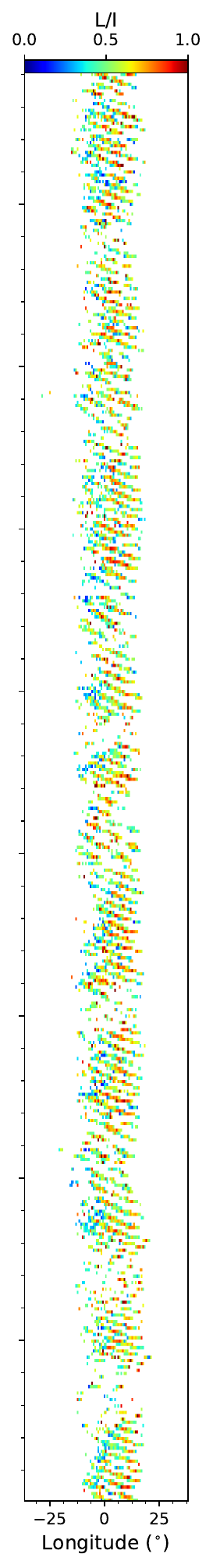}
\includegraphics[height=0.95\textheight, angle=0]{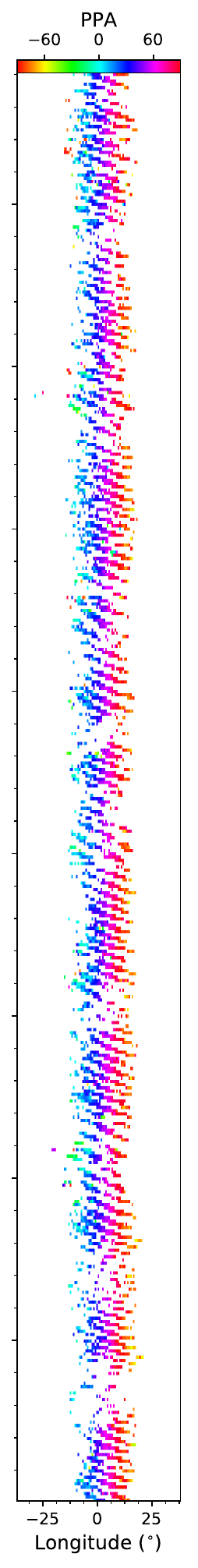}
\includegraphics[height=0.95\textheight, angle=0]{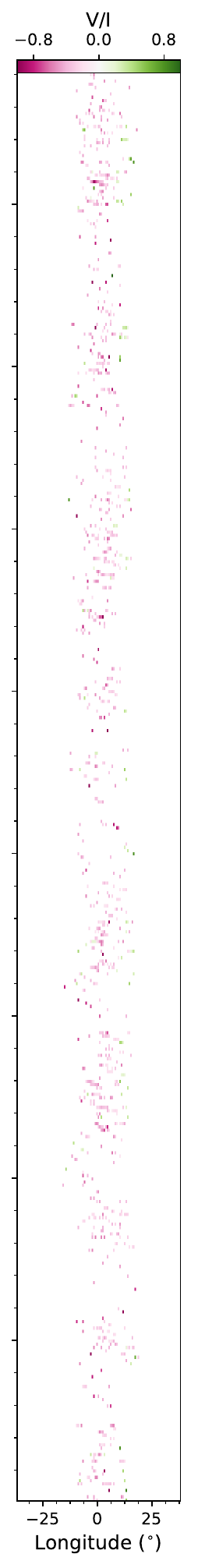}\\
\caption{Single pulse stacks of PSR J1857+0057 observed on 20220329 for I, L/I, PPA, V/I of pulses No.1-440.}
\addtocounter{figure}{-1}
\label{FigTimePhase1}
\end{figure*}


\begin{figure*}
\centering
\setlength\tabcolsep{0pt}
\begin{tabular}{llll}
\includegraphics[height=0.95\textheight, angle=0]{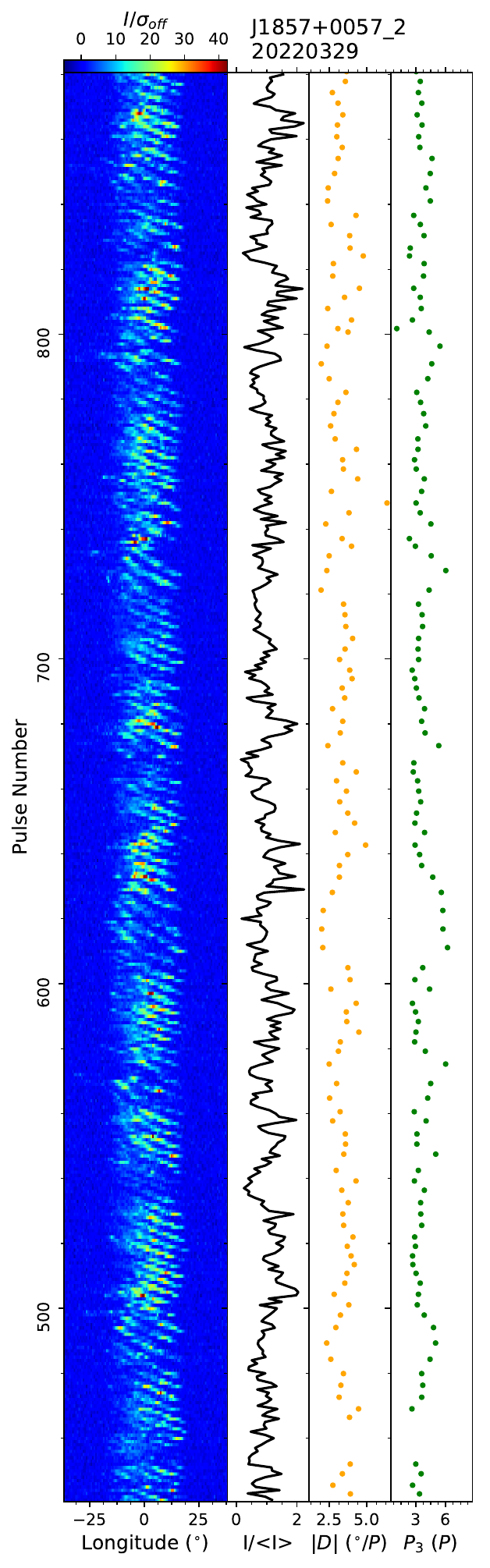}&
\includegraphics[height=0.95\textheight, angle=0]{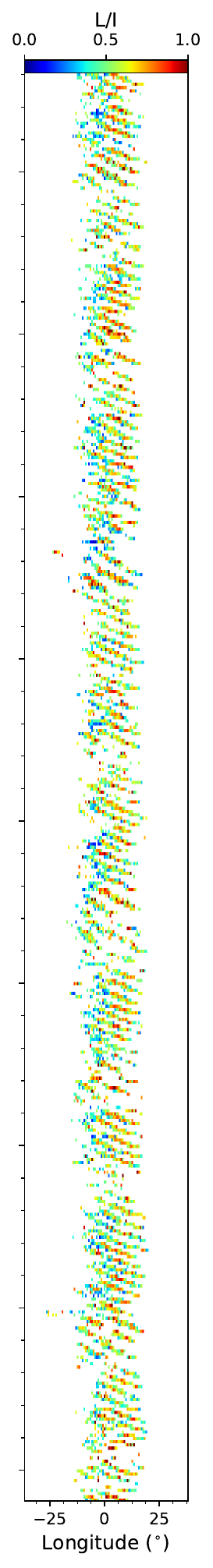}&
\includegraphics[height=0.95\textheight, angle=0]{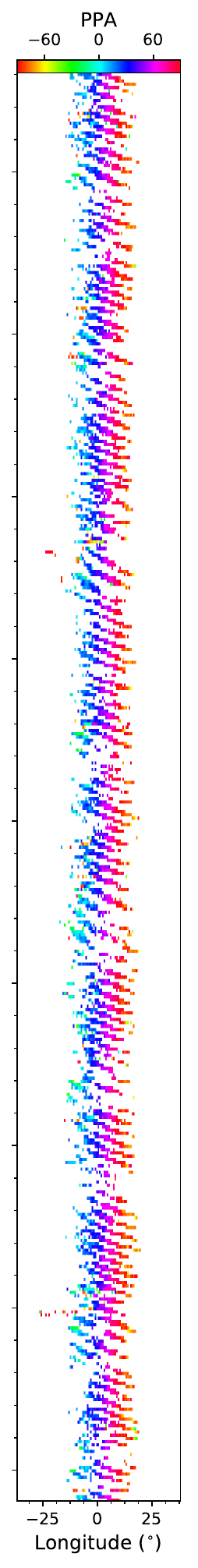}&
\includegraphics[height=0.95\textheight, angle=0]{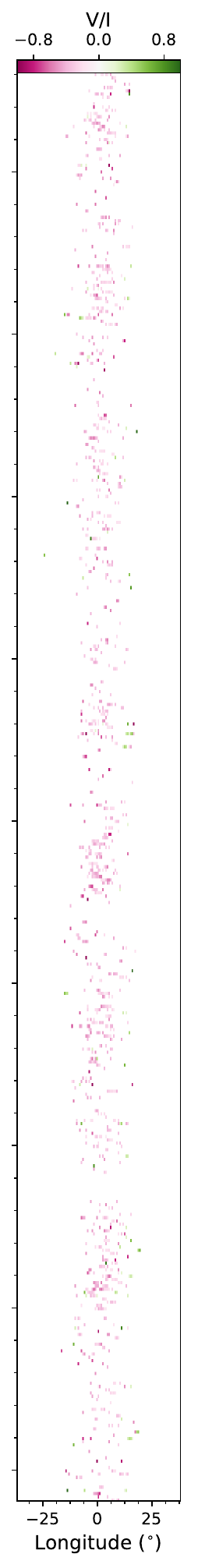}\\
\end{tabular}
\caption{Single pulse stacks of PSR J1857+0057 observed on 20220329 for I, L/I, PPA, V/I of pulses No. 441-880.}
\addtocounter{figure}{-1}
\end{figure*}


\begin{figure*}
\centering
\setlength\tabcolsep{0pt}
\begin{tabular}{llll}
\includegraphics[height=0.95\textheight, angle=0]{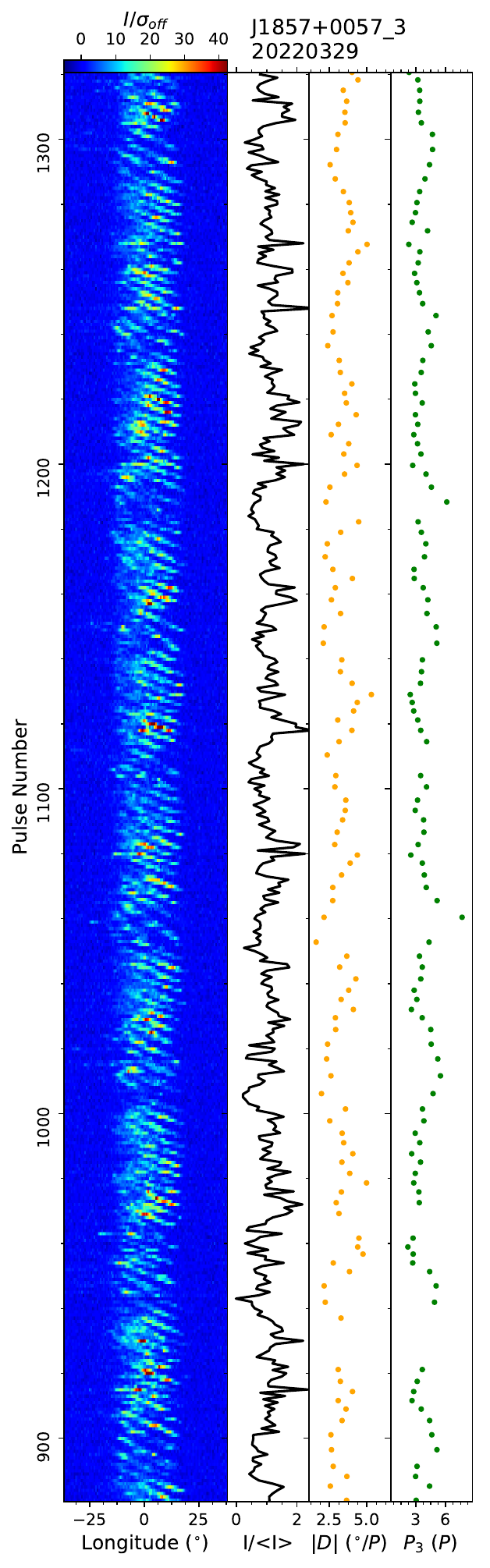}&
\includegraphics[height=0.95\textheight, angle=0]{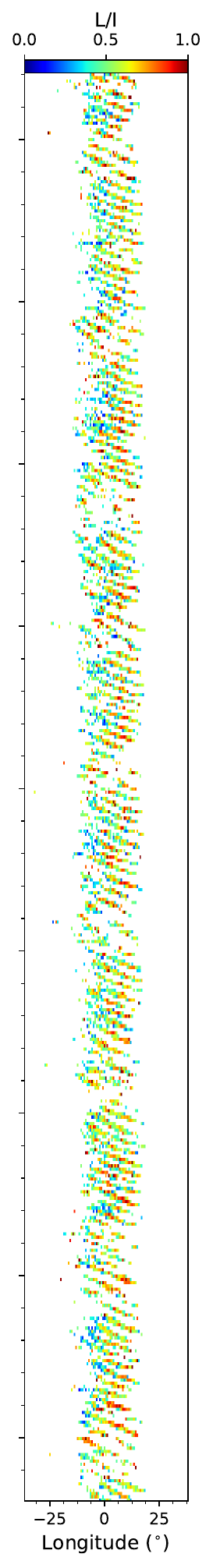}&
\includegraphics[height=0.95\textheight, angle=0]{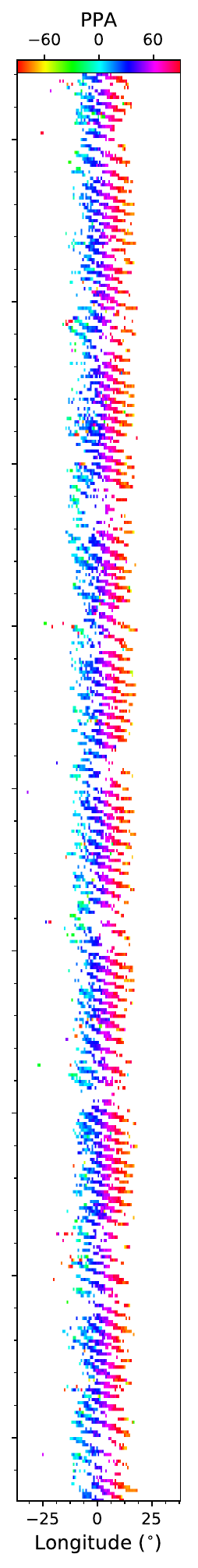}&
\includegraphics[height=0.95\textheight, angle=0]{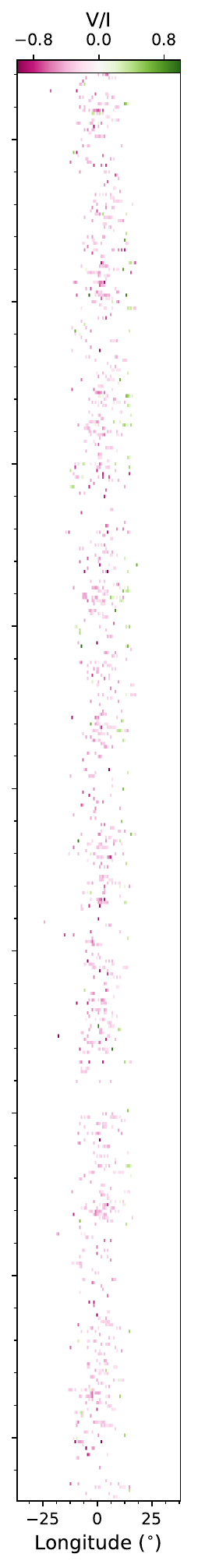}\\
\end{tabular}
\caption{Single pulse stacks of PSR J1857+0057 observed on 20220329 for I, L/I, PPA, V/I of pulses No. 881-1320.}
\addtocounter{figure}{-1}
\end{figure*}


\begin{figure*}
\centering
\setlength\tabcolsep{0pt}
\begin{tabular}{llll}
\includegraphics[height=0.95\textheight, angle=0]{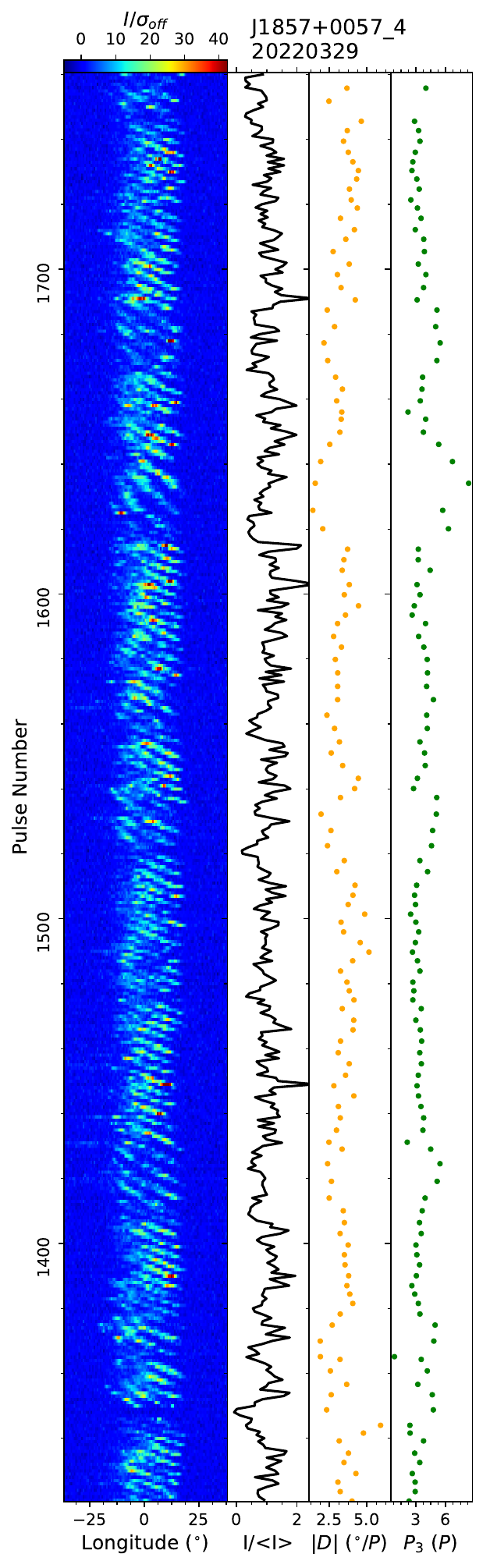}&
\includegraphics[height=0.95\textheight, angle=0]{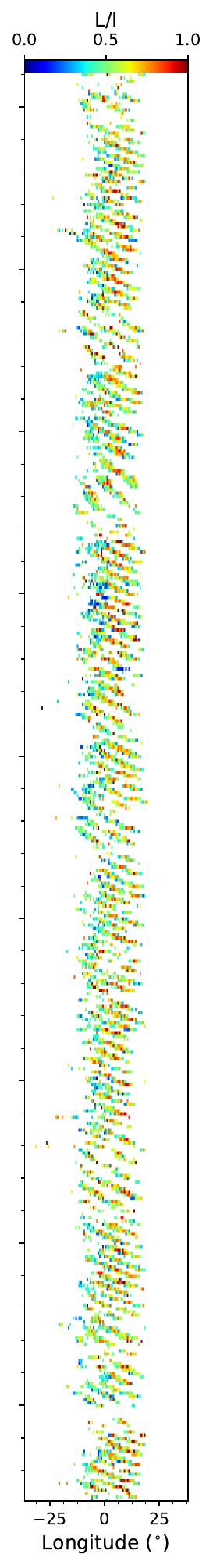}&
\includegraphics[height=0.95\textheight, angle=0]{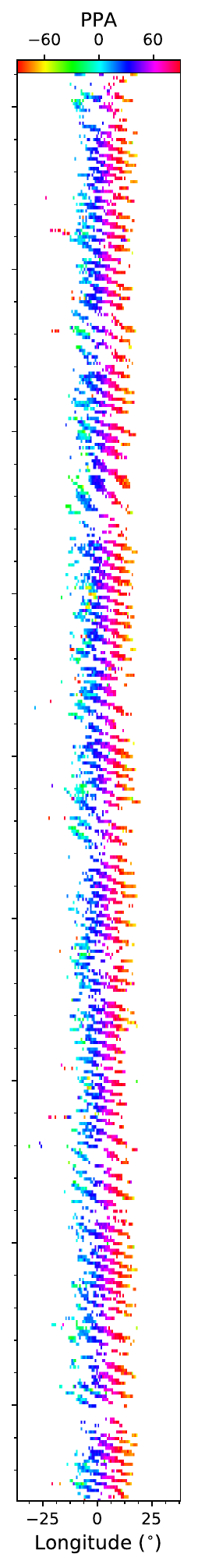}&
\includegraphics[height=0.95\textheight, angle=0]{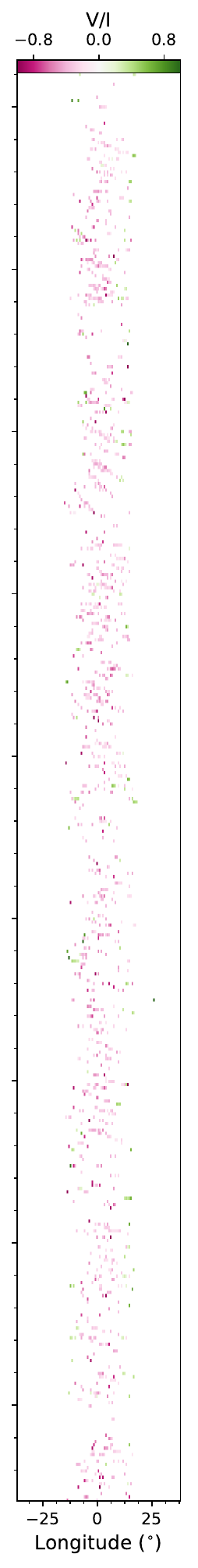}\\
\end{tabular}
\caption{Single pulse stacks of PSR J1857+0057 observed on 20220329 for I, L/I, PPA, V/I of pulses No. 1321-1760.}
\addtocounter{figure}{-1}
\end{figure*}


\begin{figure*}
\centering
\setlength\tabcolsep{0pt}
\begin{tabular}{llll}
\includegraphics[height=0.95\textheight, angle=0]{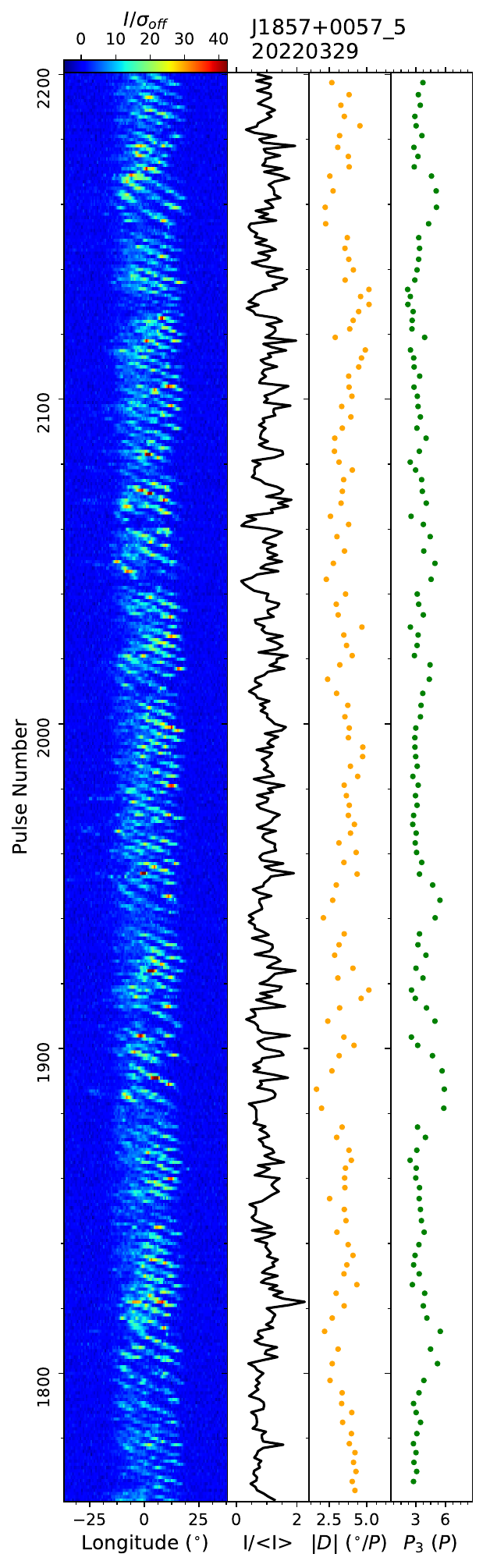}&
\includegraphics[height=0.95\textheight, angle=0]{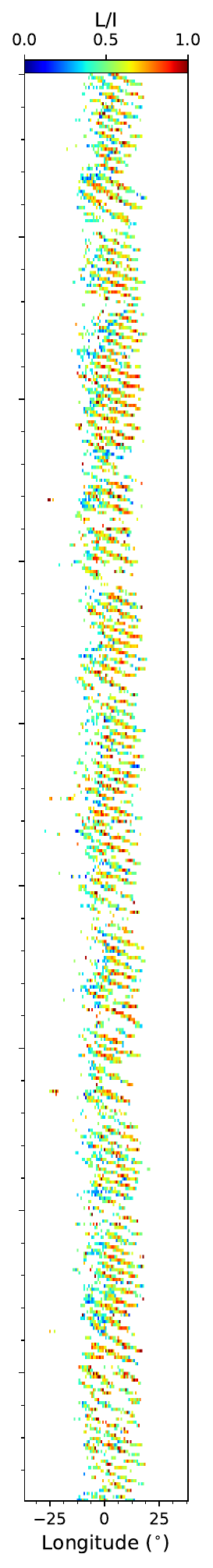}&
\includegraphics[height=0.95\textheight, angle=0]{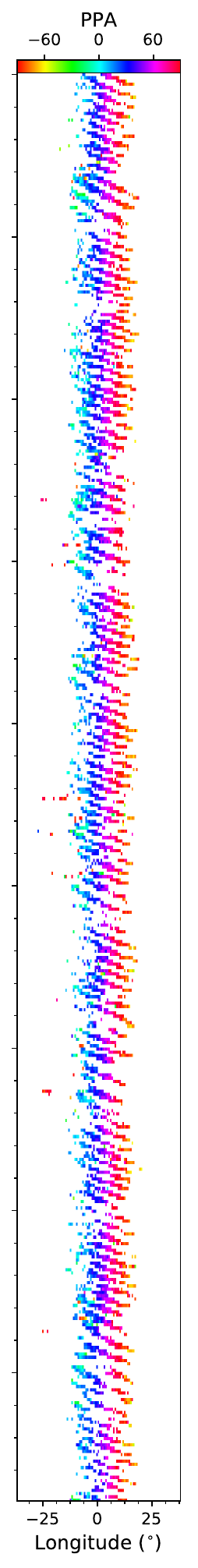}&
\includegraphics[height=0.95\textheight, angle=0]{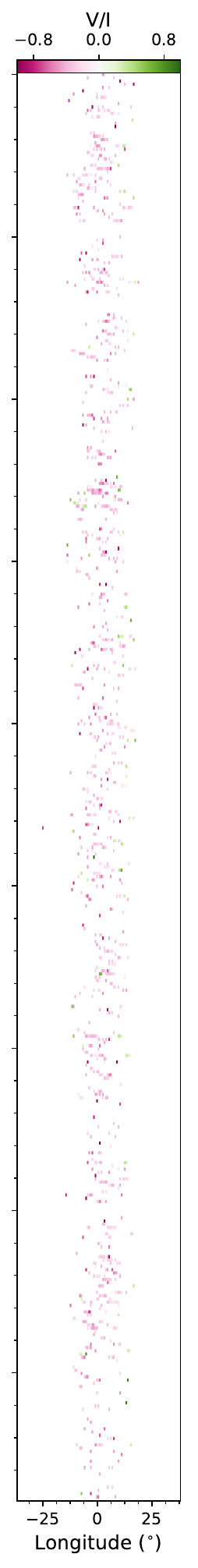}\\
\end{tabular}
\caption{Single pulse stacks of PSR J1857+0057 observed on 20220329 for I, L/I, PPA, V/I of pulses No. 1761-2200.}
\addtocounter{figure}{-1}
\end{figure*}

\begin{figure*}
\centering
\setlength\tabcolsep{0pt}
\begin{tabular}{llll}
\includegraphics[height=0.95\textheight, angle=0]{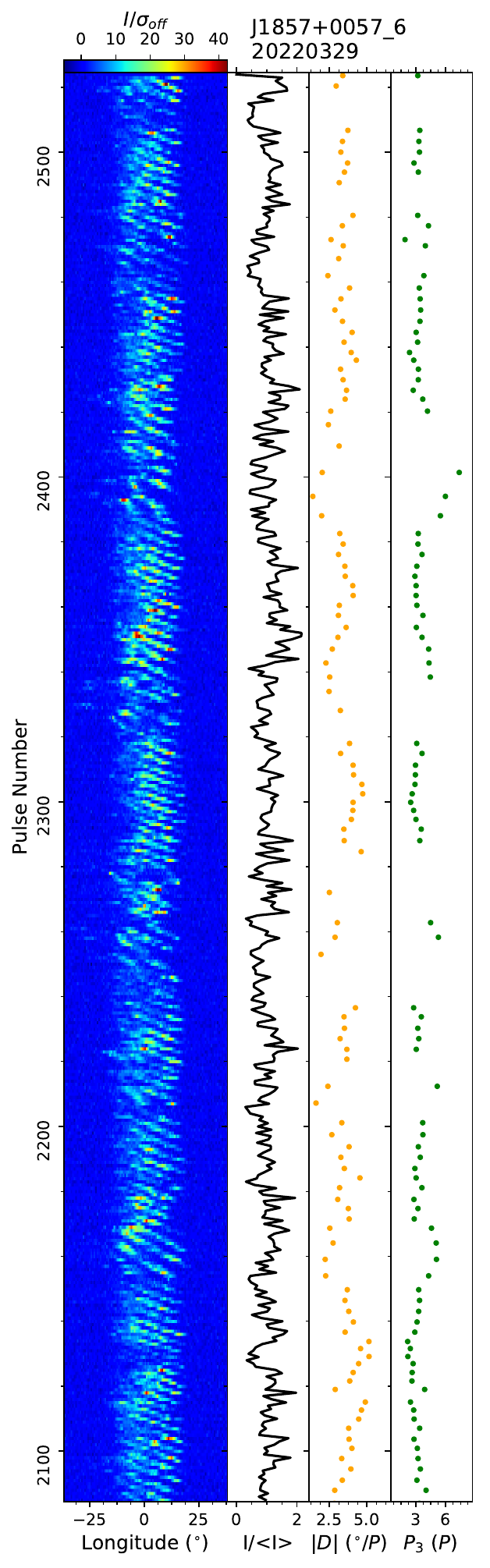}&
\includegraphics[height=0.95\textheight, angle=0]{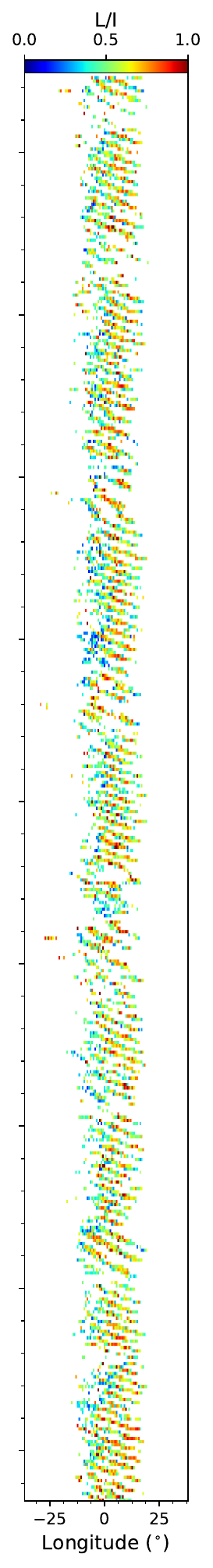}&
\includegraphics[height=0.95\textheight, angle=0]{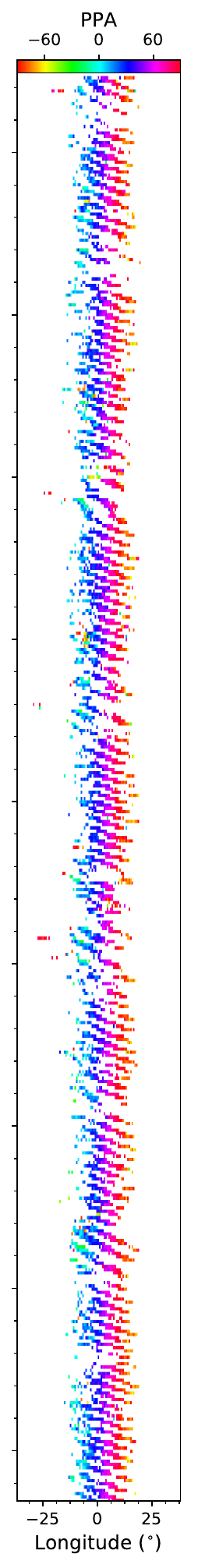}&
\includegraphics[height=0.95\textheight, angle=0]{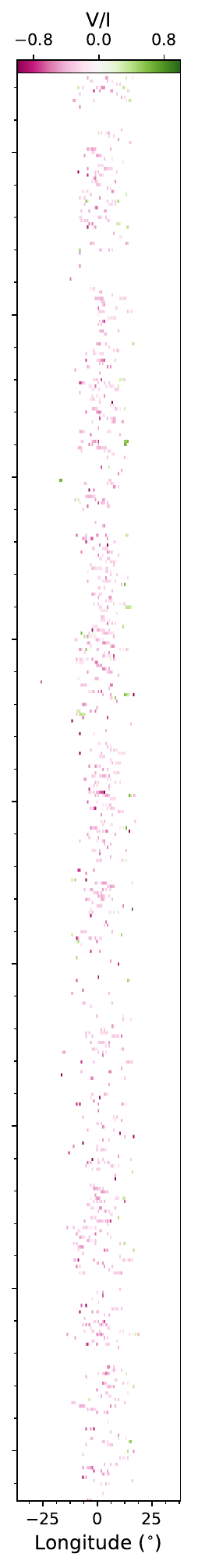}\\
\end{tabular}
\caption{Single pulse stacks of PSR J1857+0057 observed on 20220329 for I, L/I, PPA, V/I of pulses No. 2085-2524.}
\end{figure*}

\section{Determination of subpulse phases of drifting bands}

To get the drift rate of subpulses, if the drifting rate does not change over many periods for different drifting bands, one can cross-correlate consecutive pulses and directly get the phase-shift of drifting subpulses in a pair of neighbouring periods \citep{Smits2005}. However, this does not work for subpulses with a varing drifting rate. One has to determine the central phase of drifting subpulses first. A few methods have been used in literature for determination of subpulse phases of the drifting bands, namely the Gaussian-fitting to subpulses, the convolution of subpulses with a Gaussian function, or even directly using the pulse peak. The drifting rate can be determined from these phases of a train of phases.

In many cases, the drifting subpulses look like to have one component though a low signal-to-noise ratio of observations could cause subpulses apparently varing from each other. Therefore it is 
reasonable to fit a Gaussian function to get the phases of drifting 
subpulses in a drifting band. For example, \citet{van2002} assumed the subpulses to be Gaussian-shape though in reality subpulses have various shape, and then used a Levenberg-Marquardt method to locate the subpulses with high significant levels. \citet{Gajjar2017} also modeled subpulses using a Gaussian function to derive the center point. \citet{Szary2017} also used one Gaussian to fit subpulse within each component window. \citet{Zhang2019} fitted a single Gaussian to the intensity of subpulses, and the central phases of fitted Gaussians were used for deriving the drift rate of a given drifting band. In case that the subpulses of one
drifing band have a shape not resemble to a simple Gaussian component, \citet{Vivekanand1997} used several (typically two) Gaussians to fit the subpulses, and then take the sum of these fitted Gaussians as an equivalent Gaussian. If subpulses in some periods are composited of two components from two drifting bands, \citet{Szary2020} used two Gaussian functions to fit them for the phases of two drifting subpulses of the two bands.

\begin{table}
\begin{center}
\caption[]{The central phase of example drifting subpulses in Figure~\ref{FigSinPulLoc} and their uncertainty estimated by the Gaussian-fitting method and the intensity-weighted method.}
\label{TabPhaseLoc}
 \begin{tabular}{lrr}
  \hline\noalign{\smallskip}
subpulse &  Gaussian-fitting & intensity-weighted \\  
   No.      & \multicolumn{1}{c}{($^\circ$)}          & \multicolumn{1}{c}{($^\circ$)} \\
\hline
187-1     & $-$5.62  $\pm$ 0.32 & $-$5.27  $\pm$ 0.26  \\
187-2     & 6.08     $\pm$ 0.13 & 6.29     $\pm$ 0.14  \\
272-1     & $-$9.34  $\pm$ 0.13 & $-$9.77  $\pm$ 0.19  \\ 
272-2     & 0.20     $\pm$ 0.13 & 0.60     $\pm$ 0.16  \\
338-1     & $-$11.80 $\pm$ 0.09 & $-$12.19 $\pm$ 0.12  \\  
338-2     & $-$2.64  $\pm$ 0.33 & $-$2.57  $\pm$ 0.24  \\  
338-3     & 11.28    $\pm$ 0.24 & 11.19    $\pm$ 0.15  \\  
565-1     & $-$11.62 $\pm$ 0.14 & $-$11.48 $\pm$ 0.25  \\  
565-2     & $-$3.13  $\pm$ 0.21 & $-$2.56  $\pm$ 0.23  \\  
565-3     & 14.56    $\pm$ 0.85 & 14.37    $\pm$ 0.78  \\  
683-1     & $-$5.78  $\pm$ 0.70 & $-$6.36  $\pm$ 0.29  \\  
683-2     & 4.16     $\pm$ 0.14 & 3.78     $\pm$ 0.09  \\  
683-3     & 17.95    $\pm$ 0.62 & 17.38    $\pm$ 0.75  \\  
1260-1    & $-$4.06  $\pm$ 0.11 & $-$4.33  $\pm$ 0.10  \\  
1260-2    & 7.93     $\pm$ 0.14 & 7.49     $\pm$ 0.08  \\  
1669-1    & $-$8.44  $\pm$ 0.42 & $-$9.18  $\pm$ 0.79  \\  
1669-2    & 4.73     $\pm$ 0.34 & 4.86     $\pm$ 0.37  \\  
2208-1    & $-$10.98 $\pm$ 0.20 & $-$10.51 $\pm$ 0.26  \\  
2208-2    & $-$1.59  $\pm$ 0.15 & $-$1.14  $\pm$ 0.17  \\  
2208-2    & 13.13    $\pm$ 0.59 & 13.13    $\pm$ 0.76  \\  
\hline
\end{tabular}
\end{center}
\end{table}

\begin{figure*}
\centering
\setlength\tabcolsep{0pt}
\begin{tabular}{ll}
\includegraphics[height=0.22\textwidth, angle=0]{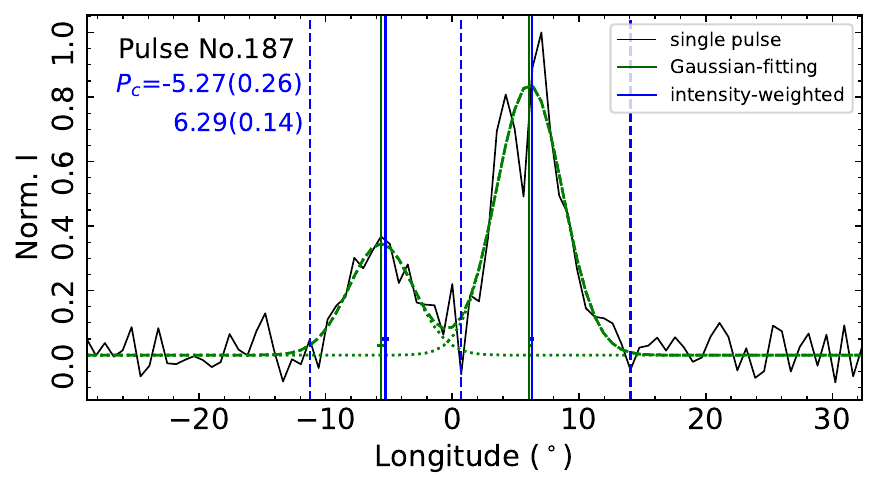}&
\includegraphics[height=0.22\textwidth, angle=0]{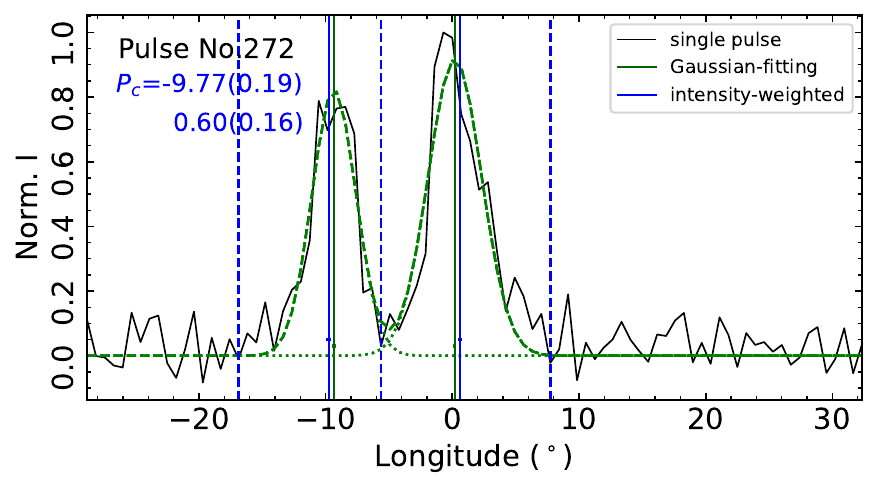}\\
\includegraphics[height=0.22\textwidth, angle=0]{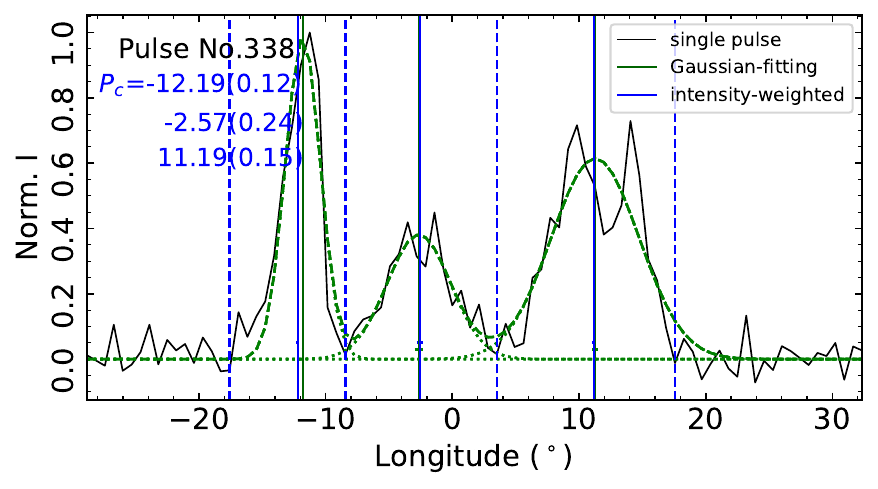}&
\includegraphics[height=0.22\textwidth, angle=0]{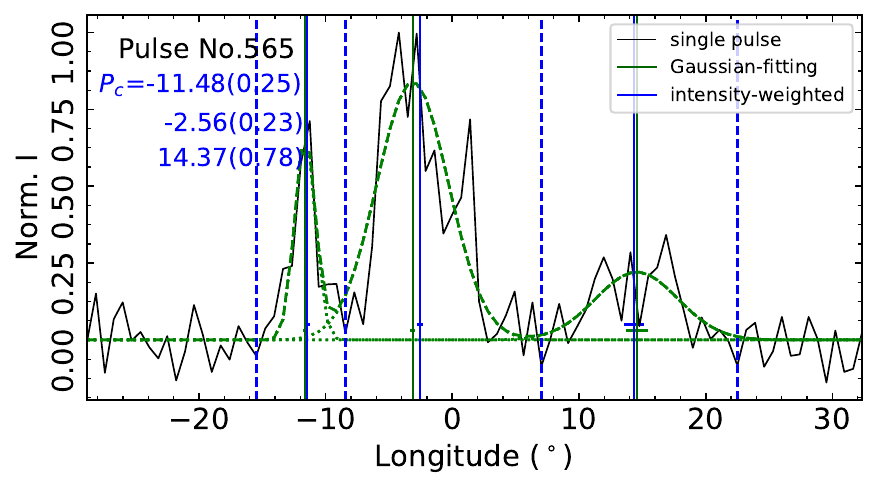}\\
\includegraphics[height=0.22\textwidth, angle=0]{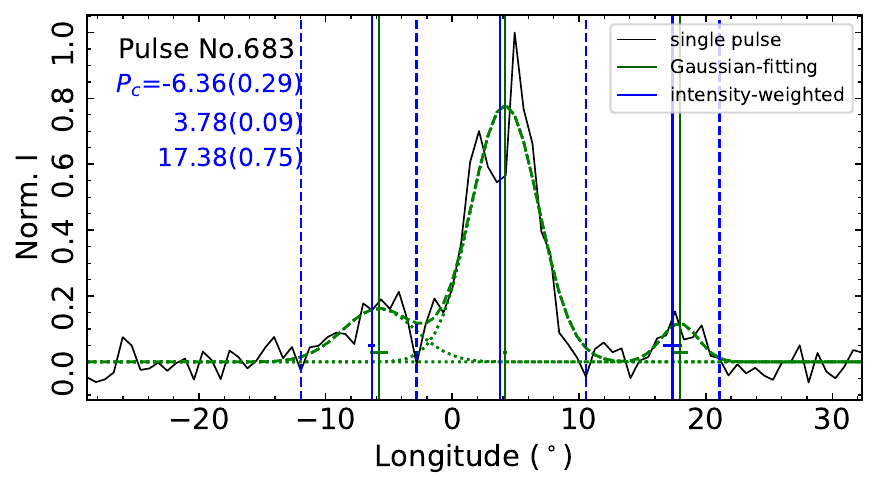}&
\includegraphics[height=0.22\textwidth, angle=0]{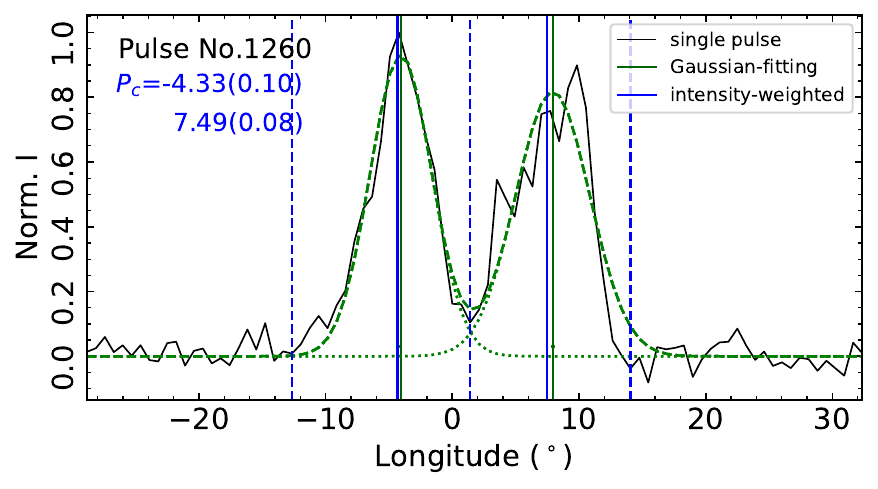}\\
\includegraphics[height=0.22\textwidth, angle=0]{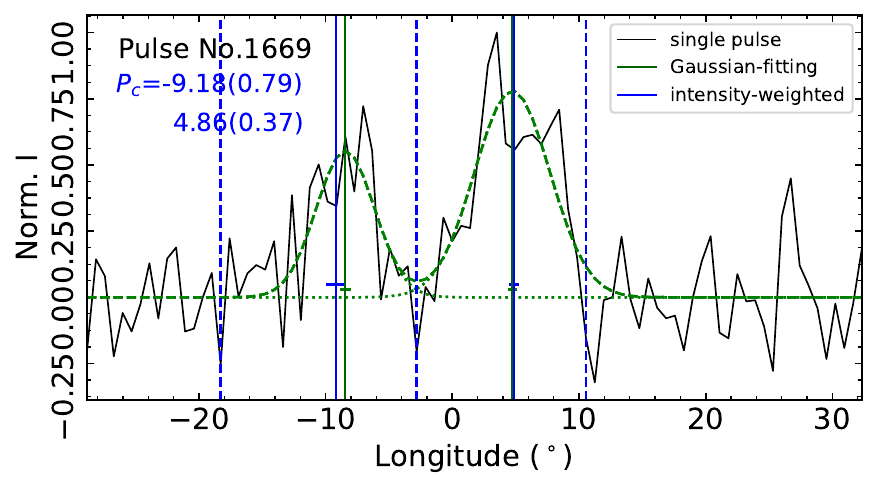}&
\includegraphics[height=0.22\textwidth, angle=0]{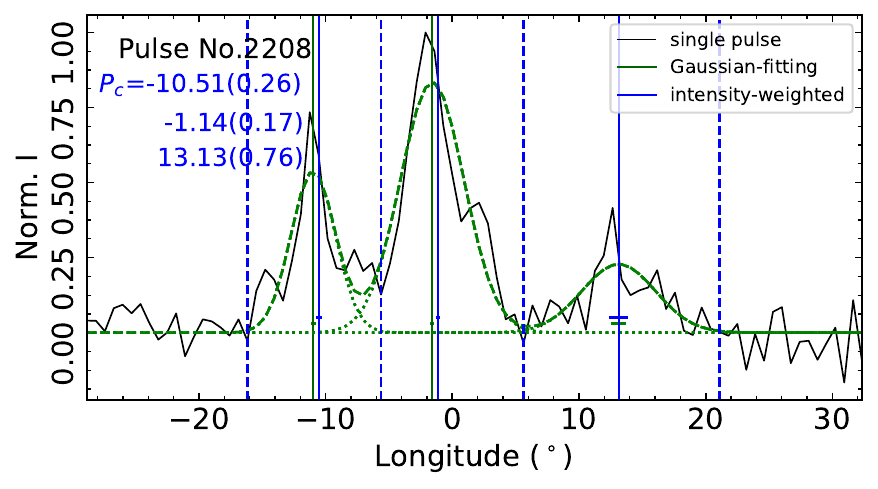}\\
\end{tabular}
\caption{Comparison of the results of central phases (solid vertical lines, with the horizontal short lines for uncertainties) of drifting subpulses estimated from the Gaussian-fitting (dashed curve in dark green) and intensity-weighted method (in blue). The vertical dashed lines indicate the boundaries of each fitted subpulse. The determined central phases are very consistent to each other.}
\label{FigSinPulLoc}
\end{figure*}

Similar to \citet{Szary2020}, we used several Gaussians to fit several subpulses in each period but of different drifting bands for PSR J1857+0057 in this paper, and obtained the phases for these subpulses for determination of drifting rate. Results for some examples are listed in Table~\ref{TabPhaseLoc}. The results for all subpulses of all periods are consistent with those in this paper, and the drifting rates and related conclusions are the same as presented in the main text of this paper. 

To smooth the noise fluctuation of observed subpulses, \citet{McSweeney2017} used a Gaussian to convolve subpulses and then took the peak phases of such smoothed subpulses as the phase of drifting  subpulses. \citet{McSweeney2022} and \citet{Szary2022} have also used such a smoothing method, and it works effectively for subpulse of well separated drifting bands.

In this paper, we develope a new method to determine the central phases of drifting subpulses with noise fluctuations due to a low signal-to-noise ratio and/or with a non-Gaussian shape. The central phase of a subpulse $P_c$ is estimated as the weighted mean of longitude $p_i$, using pulse intensity $I_i$ as weight of each bin of subpulse, 
\begin{equation}
P_c=\frac{\sum{(I_i p_i)}}{\sum{I_i}},
\end{equation}
so that the integrated energies of the subpulse on both sides are equal. The longitude boundaries of subpulses are determined by the lowest intensity between the two subpulses. In cases of no clearly defined boundary, the longitude range is roughly determined according to the average width of subpulses. The uncertainty of this subpulse central phase, $\sigma_{c}$, is given by: 
\begin{equation}
\sigma_{c}=\frac{\sigma_I}{\sum{I_i}}\sqrt{\sum(p_i-P_c)^2},
\end{equation}
here the uncertainty of the intensity $\sigma_I$ is estimated from the root-mean-square of the off-pulse ranges. Naturally the uncertainty should be smaller for a narrower pulse with a higher signal-to-noise ratio. For subpulses with a good Gaussian-shape, the central phase corresponds to the peak of the Gaussian. The values for example subpulses presented in Figure~\ref{FigSinPulLoc} are listed in Table~\ref{TabPhaseLoc}. One can see that the results are very consistent with those determined from the Gaussian fittings.




\bsp	
\label{lastpage}
\end{document}